\documentclass[a4paper,11pt]{article}
\pdfoutput=1 

\usepackage{jheppub} 

\usepackage[T1]{fontenc} 

\usepackage{amsmath}
\usepackage{amsthm}
\usepackage{amssymb}
\usepackage{braket}
\usepackage{graphicx}
\usepackage[utf8]{inputenc}
\usepackage[T1]{fontenc}
\usepackage{lmodern}
\usepackage{setspace}
\usepackage{tensor}
\usepackage{tikz-cd} 
\usepackage{genyoungtabtikz}
\usepackage{color}
\usepackage{hyperref}
\usepackage{cleveref}
\hypersetup{
    colorlinks,
    citecolor=black,
    filecolor=black,
    linkcolor=black,
    urlcolor=black
}
\usepackage{epic}
\usepackage{pict2e}
\usepackage{bbm}

\makeatletter
\newcommand{\rref}[1]{\protected@edef\@currentlabel{#1}}
\makeatother

\DeclareMathAlphabet{\mathpzc}{OT1}{pzc}{m}{it}
\DeclareMathOperator{\tr}{tr}
\DeclareMathOperator{\Id}{Id}
\DeclareMathOperator{\Perm}{\mathcal{P}}
\DeclareMathOperator{\Proj}{\mathsf{\Pi}}

\theoremstyle{definition}

 \numberwithin{equation}{section}
 \numberwithin{mdef}{section}

\let\emptyset\varnothing

\newcommand{\ZZ}{\mathbb{Z}}

\newcommand{\CC}{\mathbb{C}}
\newcommand{\lN}{\mathcal{N}}

\newcommand{\lV}{\mathcal{V}}
\newcommand{\lA}{\mathcal{A}}
\newcommand{\lB}{\mathcal{B}}

\newcommand{\gla}{\mathfrak{gl}}

\newcommand{\GL}{\mathsf{GL}}
\newcommand{\gl}{\mathfrak{gl}}
\newcommand{\wT}{\mathbb{T}}
\newcommand{\wB}{\mathbb{B}}
\newcommand{\T}{\mathbb{T}}
\newcommand{\B}{\mathbb{B}}
\newcommand{\mtwo}[4]{\left(\begin{matrix} #1&#2\\#3&#4\end{matrix}\right)}
\newcommand{\smtwo}[4]{\left(\begin{smallmatrix} #1&#2\\#3&#4\end{smallmatrix}\right)}
\newcommand{\CO}{{\mathcal{O}}}
\newcommand{\CH}{{\mathcal{H}}}
\newcommand{\BGT}{{\B^{\rm GT}}}
\newcommand{\es}{{\emptyset}}
\newcommand{\fullset}{{\bar\es}}
\newcommand{\svbas}{{\sigma}}
\newcommand{\bx}{{X}}
\newcommand{\comment}[1]{}

    \usepackage{etoolbox}
    \makeatletter
    \patchcmd{\maketitle}{\@fpheader}{}{}{}
    \makeatother

\title{Separated variables and wave functions for rational $\gl(N)$ spin chains in the companion twist frame}

\preprint{
  \begin{minipage}{0.25\linewidth}
   \flushright TCDMATH 18--15\\
    NORDITA 2018-096\\
    UUITP-49/18
  \end{minipage}
}

\author{Paul Ryan$^\alpha{}_k$}
\author{Dmytro Volin$^{\alpha}{}_{k,j}$}

\affiliation[\alpha]{School of Mathematics \& Hamilton Mathematics Institute,\\Trinity College Dublin, College Green, Dublin 2, Ireland}
\affiliation[k]{Nordita, KTH Royal Institute of Technology and Stockholm University,\\
Roslagstullsbacken 23, SE-106 91 Stockholm, Sweden}
\affiliation[j]{Department of Physics and Astronomy,\\ Uppsala University, Box 516, SE-751 20 Uppsala, Sweden}

\emailAdd{pryan@maths.tcd.ie}
\emailAdd{dmytro.volin@physics.uu.se}

\abstract{We propose a basis for rational $\gl(N)$ spin chains in an arbitrary rectangular representation $(S^A)$ that factorises the Bethe vectors into products of Slater determinants in Baxter Q-functions. This basis is constructed by repeated action of fused transfer matrices on a suitable reference state. We prove that it diagonalises the so-called B-operator, hence the operatorial roots of the latter are the separated variables. The spectrum of the separated variables is also explicitly computed and it turns out to be labelled by Gelfand-Tsetlin patterns. Our approach utilises a special choice of the spin chain twist which substantially simplifies derivations. }

\notoc
\begin{document} 

\maketitle

\flushbottom
\section{Introduction}
In quantum integrable systems, one typically has a commuting set of operators $\mathcal{B}$,  and it is often beneficial to find a separated variable (SV) representation of their common eigenvectors $|\tau\rangle$. That is to find another set of commuting operators $X_{\svbas}$ such that it has a non-degenerate spectrum, and such that the wave-functions of $|\tau\rangle$ in the co-basis of eigenvectors $\langle {\bf x}|$ of $X_{\svbas}$ factorise:
\begin{eqnarray}
\label{SoVDEF1}
\langle {\bf x}|\tau\rangle=\prod_{\svbas}\Psi_{\svbas}(x_{\svbas})\,, \quad \text{ where}\quad \langle {\bf x}|X_{\svbas}=\langle {\bf x}|x_\svbas\,.
\end{eqnarray}
One wants $X_{\svbas}$ to satisfy certain properties for them to qualify as "good" separated variables.

Firstly, the spectrum of $X_{\svbas}$ should have a reasonably regular structure. In the case we consider in this paper, we will get $x_{\svbas}=\theta_{\svbas}+\hbar\, n$, where $\theta_{\svbas}$ and $\hbar$ are fixed numbers and $n$ runs through a subset of integers. Then one can at least formally introduce ladder operators $P_{\svbas}^{\pm}$ that satisfy 
\begin{eqnarray}
\label{eq:Px}
[X_{\svbas},P_{\svbas'}^{\pm}]=\pm\hbar\, P_{\svbas'}^{\pm}\,\delta_{\svbas\svbas'}\,.
\end{eqnarray}
Finding a set of operators $P,X$ with relations \eqref{eq:Px} would correspond in classical mechanics to finding a coordinate system where the symplectic structure is canonical $\omega=\sum_{\svbas}dx_\svbas\wedge dp^{\svbas}$, and the operators $X_{\svbas}$ then correspond to the coordinates on a Lagrangian submanifold.

Secondly, the wave functions $\Psi_\svbas(x_\svbas)$ should solve a certain linear equation. It will turn out to be the Baxter equation in the case discussed in this paper. This gives one a possibility to interpret the quantum system as a quantisation of Hamilton-Jacobi equations.

Finally, there should be some procedure, intimately linked both to the separated variables and to the algebraic structure of the integrable system, that allows for a practical way of generating eigenstates of $\mathcal{B}$.

The separated variables for the case of rational $\gl(2)$ spin chains with all aforementioned properties were constructed in the works of Sklyanin \cite{10.1007/3-540-15213-X_80,Sklyanin:1991ss,Sklyanin:1995bm}. Recently there has been substantial progress towards generalising these findings to rational spin chains with higher-rank $\gl(N)$ symmetry. To summarise state of the art of this question, let us first introduce the conventions we will be using throughout the paper.

Let $\CC^N$ denote the defining representation of $\gla(N)$ and $\lV$ the finite-dimensional highest-weight representation with highest weight $\nu=[\nu_1,\nu_2,\dots,\nu_N]$. We will work with the rational $R$-matrix 
\begin{equation}
R^{\nu}(u)=u-\hbar\, \Perm^{\nu}\in {\rm End}(\CC^N\otimes \lV),
\end{equation}
where $\Perm^{\nu}$ is the generalised permutation operator  
\begin{equation}
\label{genPerm}
\Perm^{\nu}=\displaystyle\sum_{i,j=1}^NE_{ij}\otimes \pi^\nu(E_{ji})\,,
\end{equation}
with $E_{ij}$ being the $\gla(N)$ Lie algebra generators in the defining representation and $\pi^\nu(E_{ij})$ are the image of said generators in the representation $\lV$.

The physical space of the spin chain of length $L$ is $\mathcal{H}\equiv \lV^{\otimes L}$. We do not impose any inner product on it, and $\langle {\bf x}|\tau\rangle$ in \eqref{SoVDEF1} means natural pairing between elements of $\mathcal{H}^*$ and $\mathcal{H}$. We allow ourselves to refer to it as the Hilbert space, however without implying that any particular metric is used.

The bare monodromy matrix $T(u)$  is  defined as 
\begin{equation}
\label{eq:defT}
T(u)\equiv \sum_{ij}E_{ij}\otimes T_{ij}(u)=R_{L}^{\nu}(u-\theta_L)R^\nu_{L-1}(u-\theta_{L-1})\dots R^\nu_{1}(u-\theta_1)\,.
\end{equation}
It acts on $\CC^N\otimes\CH$ where the $\CC^N$ factor is referred to as the auxiliary space. The parameters $\theta_1,\dots,\theta_L$ are complex numbers called inhomogeneities. $T$ can be viewed as an $N\times N$ matrix  with entries being operators acting on $\CH$, so expressions like $T(u)T(v)$, if both $T$'s act on the same auxiliary space, stand for multiplication of these matrices.

The $RTT$ relation enjoyed by the monodromy matrix follows from the Yang-Baxter equation for the $R$-matrices and is equivalent to the commutator
\begin{equation}
(u-v)[T_{ij}(u),T_{kl}(v)]=\hbar \left(T_{kj}(u)T_{il}(v)-T_{kj}(v)T_{il}(u) \right)\,.
\end{equation}

In more abstract terms, $T_{ij}(u)$ realise a representation of the Yangian algebra $Y(\gla(N))$ which in particular is an associative algebra with the above commutation relations. The representation is uniquely defined, up to isomorphism, by the weight $\nu$ and the value of the symmetric polynomials in $\theta_a$ which appear as the coefficients in $Q_{\theta}(u)\equiv \prod_{a=1}^{L}(u-\theta_a)$. 

We will also need the twisted (dressed) monodromy matrix defined by
\begin{eqnarray}
T^{K_1,K_2}(u)=  K_1\otimes 1\times T(u) \times  K_2\otimes 1\equiv K_1TK_2\,,
\end{eqnarray}
where $K_1$ and $K_2$ are number-valued $N\times N$ matrices. They will be referred to as the left and the right twist respectively. In the case when only the right twist is present or when only the product $G=K_2K_1$ is relevant, we will use the notation $T^G$.

The Yangian contains a maximally commutative subalgebra \cite{molev2007yangians} called Bethe algebra. Consider the operator
${\rm det}(1+T^G(u)e^{-\hbar \partial_u})$ which can be expanded as \cite{Talalaev:2004qi}
\begin{equation}
\label{YangCappeli}
{\rm det}(1+T^G(u)e^{-\hbar \partial_u})=\displaystyle\sum_{a=0}^N \mathbb{T}_{a,1}(u)e^{-\hbar\,a\,\partial_u}\,
\end{equation}
where we assume the column-ordering for the determinant of a matrix with non-commutative entries i.e ${\rm det}M=\sum_{\sigma\in\mathbb{S}_N}{\rm sgn}(\sigma)M_{\sigma(1)1}M_{\sigma(2)2}\dots M_{\sigma(N)N}$. The operators $\wT_{a,1}$ are referred to as transfer matrices and they  generate the entire Bethe algebra $\mathcal{B}$. Note that the Bethe algebra is not unique but depends on the twist $G$. If we will need to emphasise the twist dependence, we will denote it by superscript, {\it e.g.} $\mathcal{B}^G$, $\wT_{a,1}^G$ {\it etc}. $\T_{a,1}$ can be equivalently defined as the transfer matrix obtained by taking the trace of the monodromy matrix with the auxiliary space in the representation $(1^a)$ of $\GL(N)$ corresponding to the Young diagram with $a$ rows each containing $1$ box. Fused transfer matrices corresponding to generic finite-dimensional representations $\lambda=[\lambda_1,\lambda_2,\ldots,\lambda_N]$ will be denoted $\T_\lambda$. In the special case when the Young diagram is rectangular with $a$ rows and $s$ columns we denote the corresponding transfer matrix $\T_{a,s}$. The transfer matrices $\T_\emptyset=\T_{a,0}=\T_{0,s}$ corresponding to the empty diagram are simply the identity operators. 

The separation of variables (SoV) program is to construct separated variable representations for eigenstates of the Bethe algebra $\mathcal{B}$. Constructing just some commuting set of $X$'s that factorise the Hilbert space can be actually very straightforward. Indeed, the Hilbert space is $\lV^{\otimes L}$ and hence it is already factorised in the natural $\GL(N)$-covariant basis. We can hence elect Cartan generators of $\gl(N)$ acting on the spin chain nodes as separated variables, i.e. for $\svbas\equiv (\alpha,i)$, define $X_{\alpha,i}=1\otimes 1\otimes\ldots \pi^{\nu}(E_{ii})\otimes\ldots \otimes 1$, where $\pi^{\nu}(E_{ii})$ acts on the $\alpha$-th position. This definition has several problems though. The first one is that the spectrum of such algebra is degenerate, unless $\pi^{\nu}$ is of special type. It is overcome by extending the Cartan subalgebra to the Gelfand-Tsetlin (GT) algebra which is a  canonically constructed commutative subalgebra of  $U(\gl(N))$. It has, in contrast to the Cartan subalgebra, non-degenerate spectrum. The second problem is that $X$'s are not immediately linked to $T_{ij}$, the generators of the Yangian algebra. The latter should be thought as our "observables" defining the physical realm, hence any physically relevant operator should be related to $T_{ij}$. This difficulty is resolvable as well, by using the generalisation of the GT basis for the case of Yangian, and we  briefly cover this known \cite{molev2007yangians} construction in section~\ref{sec:Spectrum}.

A more serious issue is that it is not obvious whether a given basis that factorises the Hilbert space also factorises the Bethe vectors $\ket{\tau}$ we are interested in. For instance, the GT basis does not perform the job, at least in the approach we will discuss in this paper. Still, knowing the structure of the GT algebra turns out to be very useful because, as we will see, there is a natural basis in the GT algebra isospectral to the good SV basis that shall be constructed. 

An important step forward towards finding a good SV basis was made in the recent paper by Maillet and Nicolli \cite{Maillet:2018bim} where they made a very simple but powerful observation. If $|\tau\rangle$ is an eigenvector of $\mathbb{T}\equiv \wT_{1,1}$ then the following scalar product naturally factorises: 
\begin{eqnarray}
\langle 0 |\prod_{\theta \in \Lambda}\wT(\theta)|
\tau\rangle=\prod_{\theta\in \Lambda}\wT^{\tau}(\theta) \langle 0 |\tau\rangle\,,
\end{eqnarray}
where $\Lambda$ is a certain set of rapidities and $\bra{0}$ is some fixed reference state, and $\wT^{\tau}$ is the eigenvalue of the transfer matrix. Hence it is tempting to define $X$'s as the operators that have  $\langle \Lambda|\equiv \langle 0 |\prod\limits_{\theta \in \Lambda}\wT(\theta)$ as their left eigenvectors. A rather non-trivial technical question is whether we can find such regular enough collection of sets $\Lambda$ that vectors $\langle \Lambda |$ form a co-basis. This was positively resolved in \cite{Maillet:2018bim}. An important limitation of the proposed idea is that, while it poses a promising SoV approach, construction of the $\bra{\Lambda}$-basis does not  suggest how to construct the eigenstates of the Bethe algebra. Indeed, Maillet and Nicolli {\it de-facto} define the operator 
\begin{eqnarray}
\label{Bdef}
\mathbb{B}(u)=\prod_{\svbas}(u-X_{\svbas})\,,
\end{eqnarray}
through their separated variables $X_{\svbas}$. It is not difficult to show that 
\begin{eqnarray}
\label{eq:eigstates}
|\tau\rangle=\prod_{r}\B(u_r)|\Omega\rangle,
\end{eqnarray}
where $u_r$ solve the Bethe equations and $\ket{\Omega}$ is an appropriately chosen state, are indeed eigenstates of the Bethe algebra. An alternate way of writing this (up to an overall sign) is 
\begin{equation}
|\tau\rangle=\prod_{\svbas}Q(X_\svbas)|\Omega\rangle,
\label{eq:QXb}
\end{equation}
where $Q$ are momentum-carrying Baxter $Q$-functions. This construction of a separated variable basis works well for the defining representation, however $X_{\svbas}$ are not directly expressed in terms of Yangian generators, and therefore  \eqref{Bdef} used as the definition of $\B$ has limited practical usage. For instance, \eqref{eq:eigstates} can be understood only as a rewriting of \eqref{eq:QXb}. Furthermore, it is not immediately clear what modifications are needed for the case of higher representations. 

A complementary and much older idea is to seek for a special operator $\mathbb{B}(u)$ built directly from the Yangian generators with the goal to claim its operatorial zeros as separated variables. In this case one should find the spectrum of $\mathbb{B}(u)$ and prove that indeed \eqref{eq:eigstates} generates the eigenstates, which can be a non-trivial computation.

In \cite{Gromov:2016itr}, Gromov, Levkovich-Maslyuk, and Sizov proposed the following operator
\begin{eqnarray}
\label{Bgooddef}
\mathbb{B}(u)\propto \sum_{k_1,k_2,\ldots,k_{N-2}}{\bf T}\left[^{k_1}_{N}\right]{\bf T}^{[2]}[^{k_2}_{k_1\,,{N}}]{\bf T}^{[4]}\left[^{k_3}_{k_2\,,{N}}\right]\ldots{\bf T}^{[2N-4]}\left[^{1\ldots N-1}_{k_{N-2}\,{N}}\right]\,,
\end{eqnarray}
where  $k_i=k_i^{1}k_i^{2}\ldots k_i^{i}$ is a multi-index with $1\leq k_i^{1}< k_i^{2}<\ldots< k_i^{i}\leq N-1$. The quantum minors in the construction are defined as
\begin{equation}
\label{qminor}
{\bf T}\left[^{i_1\dots i_m}_{j_1\dots j_m}\right]=\sum_{\sigma\in\mathbb{S}_m}{\rm sgn}(\sigma){\bf T}_{i_{\sigma(1)}j_1}{\bf T}^{[-2]}_{i_{\sigma(2)}j_2}\dots {\bf T}^{[2-2m]}_{i_{\sigma(m)}j_{m}}\,,\quad {\bf T} \equiv T^{K_1,K_2}\,,
\end{equation}
and $f^{[2n]}(u)\equiv f(u+\hbar\,n)$ is the notation for the shift of the spectral parameter $u$. Note that $\mathbb{B}$ depends on both $K_1,K_2$ and it is assumed that $K_1,K_2$ are such that $\mathbb{B}$ is not nilpotent. In particular $K_1,K_2$ can't be both diagonal.

The sign $\propto$, here and elsewhere in the paper, means that the equality holds up to an overall normalisation. In the case when such equalities serve to define an object, its normalisation should be stated separately. For instance, \eqref{Bgooddef} defines $\B(u)$ and we additionally agree to always normalise $\B(u)$ in such a way that it is a monic polynomial in $u$.

The proposal \eqref{Bgooddef} generalises, in a slightly modified form, Sklyanin's approach for $N=3$ \cite{Sklyanin:1992sm} to higher values of $N$. Sklyanin's ideas in turn rely on analogy with the classical approach \cite{Sklyanin:1992eu} where separated variables, now zeroes of $\mathbb{B}$, would label the dynamic divisor, or equivalently the properly normalized Baker-Akhiezer function, which is an eigenvector of the classical monodromy matrix. 

Based on experimental evidence for small $N$ and $L$, the authors of \cite{Gromov:2016itr}  conjectured  the spectrum of $\mathbb{B}(u)$ and also conjectured that \eqref{eq:eigstates} are indeed transfer matrix eigenstates. The latter was proven in \cite{Liashyk:2018qfc} for $N=3$ and symmetric powers of the defining representation, however the proof is quite tedious and difficult for higher rank generalisations.

Maillet and Nicolli conjectured in \cite{Maillet:2018bim} that their separated variables coincide with operatorial zeros of $\mathbb{B}$ defined by \eqref{Bgooddef} and checked for $N=3$ and small $L$.

In this paper, we propose a proof of the stated conjectures for arbitrary $N$ and $L$ and, furthermore,  extend the study to a more general class of representations $\lV$ which provides new insights into the nature of the problem. 

The paper  consists of two parts. In section 2 we explore connections between $\mathbb{B}$ and the GT algebra, which allows us to explicitly compute the spectrum of $\mathbb{B}$. This part is done for arbitrary representation  $\lV$. 

In section 3 we introduce the separated variables  by generalisation of the approach in \cite{Maillet:2018bim}  and show that they are operatorial zeros of $\mathbb{B}$. This construction is designed for arbitrary representations of type $\lV=(S^A)$ which are called rectangular representations, alleging to the shape of the corresponding Young diagram - the highest weight is $[S,S,\ldots,S,0,\ldots,0]$, where $S$ appears $A$ times and $0$ appears $N-A$ times. For arbitrary values of $A,S$, we derive the explicit expression for the wave functions \eqref{SoVDEF1} in terms of Slater determinants in Baxter Q-functions. As outlined in the conclusions, restriction of this result for the case $A=1$ implies  \eqref{eq:eigstates} and \eqref{eq:QXb} which are  two not immediately related statements when $\mathbb{B}$ is defined by \eqref{Bgooddef} but not by \eqref{Bdef}.

It is very beneficial to select a reference frame where the twist matrices become of a very special form. Namely, we propose to set $K_1$ to the identity matrix and $K_2=G$ to the so-called (dual) companion matrix for eigenvalues $z_1,z_2,\ldots\ldots,z_N$:
\begin{equation}
\label{eq:Gcompanion}
G=\left(
\begin{array}{cccccc}
\chi_1 & -\chi_2 & \chi_3 & \dots & (-1)^{N-2}\chi_{N-1} & (-1)^{N-1}\chi_N \\
1 & 0 & 0 & \dots & 0 &  0 \\
0 & 1 & 0 & \dots & 0 &  0 \\
\vdots & \vdots & \vdots & \ddots & \vdots & \vdots \\
0 & 0 & 0 & \dots & 1 & 0
\end{array}
\right)\,,
\end{equation}
where $\chi_a$ are the symmetric polynomials of $z$'s defined by $\prod_{i=1}^N (t+z_i)=\sum_{i=0}^N t^{N-i}\chi_i\,.$ In this specific companion twist frame computations simplify significantly allowing to explicitly derive the aforementioned results. In particular, $\mathbb{B}$ stops depending on $z$'s and the only twist-dependent quantities are the $Q$-functions, as solutions of twist-dependent Bethe equations, and the physical vacuum $|\Omega\rangle$.

\section{Spectrum of $\B$ \label{sec:Spectrum}}
In this section we will find the spectrum of $\B$ defined by \eqref{Bgooddef}. First let us make some comments to set up the framework. 

$\B$ is a polynomial in $T_{ij}^{K_1,K_2}$ and it rather non-trivially depends on both $K_1,K_2$. Its action $\prod_{r}\B(u_r)|\Omega\rangle$ on a specially selected state $|\Omega\rangle$ is conjectured in general and  proven in certain cases to produce the eigenstates of the Bethe algebra $\mathcal{B}^g$, $g=K_2K_1$. Hence there is a family of different choices of $K_1,K_2$, thus a family $\B$'s, that should accomplish the same goal. This curious property was remarked in \cite{Gromov:2016itr} and remains unproven save for some simple enough cases. Note that it non-trivially utilises that $u_r$ satisfy Bethe equations.

If $g$ is diagonal, it is tempting to take $K_1,K_2$ diagonal as well. However this choice leads to a nilpotent $\B$ which renders it impossible to use operatorial zeros of $\B(u)$ as separated variables (on the other hand if $A=1$ a nilpotent $\B$ is still expected to generate eigenstates of the Bethe algebra via \eqref{eq:eigstates},  as one can learn from explicit examples for $N=2,3$ cases). Hence a different set up should be considered.

If one is to follow the original ideas of Sklyanin then one does not need to use both left and right twist. Instead one can put, say, $K_1=1$ and and $K_2=g$, where $g$ is non-diagonal. Such an option was also considered in \cite{Kazama:2013rya} with the aim of restoring no twist at all (i.e. $K_1$ and $K_2$ are the identity) at the end of computations as a special limit.

Another option which is used in \cite{Gromov:2016itr} is to perform a similarity transformation  on the auxiliary space 
\begin{equation}
T(u)g\rightarrow KT(u)gK^{-1}
\end{equation}
which does not affect the Bethe algebra $\mathcal{B}^g$. In other words, we choose $K_1=K$ and $K_2=gK^{-1}$. In this way one can keep $g$ diagonal and yet get a non-nilpotent $\B$.

We  perform one more step now and perform the same similarity transformation in the physical space
\begin{equation}
\label{symphys}
\begin{split}
KT(u)gK^{-1} & \rightarrow  \Pi(K)KT(u)gK^{-1}\Pi(K)^{-1}\\
& =K\Pi(K)T(u)\Pi(K)^{-1}gK^{-1}\\
& =T(u)KgK^{-1}\,,
\end{split}
\end{equation}
where $\Pi(K)$ denotes the appropriate representation of the $\GL(N)$ group element $K$ acting on the physical space, and where we used the $\GL(N)$ invariance of the monodromy matrix in the second equality. The net effect of this transformation is in cancelling off the left twist $K_1$.

The structure $TKgK^{-1}$ will be the starting point for us. In the spirit of Sklyanin's approach, we  consider  the case of only the right twist $K_2=G=KgK^{-1}$ present, and explicitly construct it to be non-diagonal. We  will focus on designing such a $G$ such that the computations simplify substantially, with the final conclusion being that the choice \eqref{eq:Gcompanion} of the companion matrix is a convenient one. Note that $\B\equiv\B^{1,G}$  immediately serves to diagonalise $\mathcal{B}^G$. To diagonalise the original Bethe algebra $\mathcal{B}^g$ it suffices to undo the similarity transformation \eqref{symphys} of the physical space. Obviously, this operation won't affect any of the properties we derive, however it will change $\B$ to $\B=\B^{K_1,K_2}$ -- a more intricate combination of bare $T$'s with $K_1=K$ and $K_2=gK^{-1}$, i.e. the one used in \cite{Gromov:2016itr}.

In the outlined strategy we derive the requested properties of $\B$ without  performing the similarity transformations in the auxiliary space. While, by conjecture, performing such generic enough transformations won't spoil the properties of $\B$, it is not proven and we want to avoid this step.

As we can always return to the frame where $g$ is  any  matrix similar to $G$, in particular the diagonal one if $z_i$ are pair-wise distinct, from now on we will focus on working only with the twist $G$ and diagonalisation of the Bethe algebra $\mathcal{B}^G$.
\subsection{Revisiting the $N=2$ case}
For $\gl(2)$ spin chains it is common to represent the dressed monodromy matrix as $T^G=\left(\begin{smallmatrix}A^G&B^G\\C^G&D^G\end{smallmatrix}\right)$, and then $\B= B^G$. This case has been studied in great detail, see for example \cite{Sklyanin:1991ss,Sklyanin:1995bm,Kazama:2013rya}. Let us take a slightly different point of view now and rephrase the definition of $\B$ as follows
\begin{eqnarray}
\label{BT}
\B \propto \tr\left[ T^G \left(\begin{matrix}0&0\\1&0\end{matrix}\right)\right]=\wT^{H}\,,\quad H=G\left(\begin{matrix}0&0\\1&0\end{matrix}\right)\,,
\end{eqnarray}
that is we can interpret $\B$ as a transfer matrix of the spin chain with the singular twist $H$. One has $\det H=0$ and we require $\tr H\neq 0$ in order to get a non-nilpotent $\B$. Apart from that, $H$ can be fairly arbitrary so let us make some simple choice and put
\begin{eqnarray}
\label{eq:BT11}
\B= T_{11}\,,
\end{eqnarray}
where $T_{11}$ is an element of the bare monodromy matrix.

Let us find $G$ providing this choice. For $G=\smtwo \alpha\beta\gamma\delta$, $H=\smtwo{\beta}{0}{\delta}{0}$. We should set $\delta=0$ to get \eqref{eq:BT11} and, as the normalisation is always adjustable, we will set $\gamma=1$. After specifying that $z_1,z_2$ are the eigenvalues of $G$, the latter is now uniquely fixed to be
\begin{eqnarray}
\label{eq:comptwist2}
G=\mtwo {z_1+z_2}{-z_1z_2}{1}{0} \,.
\end{eqnarray}
Hence we can set up the SoV program for the Bethe algebra $\mathcal{B}^G$ with $\B=T_{11}$.

The outlined ideas already exist in the literature, albeit with a different emphasis and motivation. Relations of type \eqref{BT} can be found in \cite{belliard2018modified,Maillet:2018bim}. The separated variables program for the twist matrix \eqref{eq:comptwist2} in the case $z_1=-z_2=1$ completely on the left and $\B=T_{22}$ was performed in \cite{Niccoli:2012vq}. We hence won't discuss the details here in full but only mention a way to derive the spectrum of $\B$ as it is useful for further generalisations. 

From the definition \eqref{eq:defT} it is easy to derive that
\begin{eqnarray}
\label{eq:T11spec}
T_{11}(u)=\prod_{\alpha=1}^L (u-\theta_\alpha-\hbar\ \pi^{\nu}_\alpha(E_{11}))+{\rm Nil_L}\,,
\end{eqnarray}
where $\pi^{\nu}_\alpha(E_{11})$ is the $E_{11}$ generator acting on the $\alpha$'th site of the spin chain. The term ${\rm Nil_L}$ designates a nilpotent operator. Its function is to permute, with some coefficients, states at different sites of spin chain in a way that the spin chain position of lower-weight states is shifted left. Hence there exist a basis where ${\rm Nil_L}$ is upper-triangular while the first term of \eqref{eq:T11spec} is diagonal. Then, given that the first term is non-degenerate, ${\rm Nil_L}$ does not affect the eigenvalues of $\B$. So for $\B=\prod\limits_{\alpha=1}^L(u-X_\alpha)$, the spectrum of separated variables $X_\alpha$ is found to be $X_\alpha=\theta_\alpha+\hbar\,n$, where $n$ is an integer in the range $\nu_2\leq n\leq \nu_1$.
\subsection{Gelfand-Tsetlin algebra}
We just saw that, for $N=2$, $\B$ can be interpreted as a transfer matrix in the singular twist limit. Therefore one can say that separated variables of the Bethe algebra $\mathcal{B}^G$ are generated by the Bethe algebra $\mathcal{B}^H$, and the spectrum of the transfer matrix $\wT^{H}$ generating $\mathcal{B}^H$ is particularly easy to compute.

What about attempting to generalise this observation to higher rank? By singular twist we mean the matrix $H={\rm diag}(y_1,y_2,\ldots,y_N)$ with subsequent limit $y_1\gg y_2\gg \ldots \gg y_N$. In this subsection we discuss the properties of transfer matrices in this limit. One can observe that they become simply quantum minors
\begin{eqnarray}
\label{eq:qminors}
\wT_{a,1}^H \propto T\!\left[^{12\ldots a}_{12\ldots a}\right]\,.
\end{eqnarray}
These minors were defined in \eqref{qminor},  note however that now we are operating with bare monodromy matrices.

The minors \eqref{eq:qminors} are special. They are quantum determinants of $Y(\gla(a))\subset Y(\gla(N))$, and they are diagonal in the Galfand-Tsetlin basis of $Y(\gla(N))$, moreover their common spectrum is non-degenerate. In other words the Bethe algebra in the singular twist limit actually coincides with the Gelfand-Tsetlin algebra.

We now review the construction of the Gelfand-Tsetlin basis. A full account and proofs can be found in {\it e.g. } \cite{molev2007yangians,molevgelfand}, see also \cite{nla.cat-vn1878494}. Note that GT algebras were already considered as candidates for separated variables in \cite{Valinevich:2016cwq}.

\paragraph{GT basis for $\gla(N)$.}
Consider a finite-dimensional representation $\lV_{\lambda}$ of $\gla(N)$ with highest-weight $\lambda=[\lambda_1,\lambda_2,\dots,\lambda_N]$. There exists a unique up to normalisation vector called the highest-weight state $\ket{\text{HWS}}$ which satisfies
\begin{equation}
\begin{split}
& E_{ii}\ket{\text{HWS}}=\lambda_i \ket{\text{HWS}}\,, \\
& E_{ij}\ket{\text{HWS}}=0, \ i<j\,.
\end{split}
\end{equation}
$\gla(N)$ has a subalgebra naturally identified with $\gla(N-1)$:
\begin{equation}
\gla(N-1)={\rm span}\{ E_{ij},\ i,j=1,\dots,N-1\}\,.
\end{equation}
The restriction of $\lV_\lambda$ to $\gla(N-1)$ decomposes into a direct sum of $\gla(N-1)$ irreps $\lV_{\lambda'}$ with weights $\lambda'=[\lambda'_1,\dots,\lambda'_{N-1}]$ that satisfy the branching rules
\begin{equation}
\label{branchingrules}
\lambda_i\geq \lambda'_i\geq \lambda_{i+1},\ i=1,\dots,N-1\,.
\end{equation}
Each weight that satisfies this rule appears exactly once in the direct sum.

The process can be repeated by looking at the chain of subalgebras
\begin{equation}
\gla(1)\subset \gla(2)\subset\dots\subset\gla(N-1)\subset\gla(N)\,.
\end{equation}
Let $[\lambda_{k1},\lambda_{k2},\dots,\lambda_{kk}]$ denote the highest weight of the irrrep of the $\gla(k)$ algebra appearing in the above chain. Then we construct the following array $\Lambda$
\begin{equation}
\begin{array}{ccccccccc}
\lambda_{N1} &  & \lambda_{N2} &  & \dots & \ & \lambda_{N,N-1} &  & \lambda_{NN} \\
 & \lambda_{N-1,1} &  & \lambda_{N-1,2} & \dots & \lambda_{N-1,N-2}&  \ & \lambda_{N-1,N-1} & \ \\
 & \ & \dots & \ & \dots & \ & \dots &  \\
 & \ & \ & \lambda_{21} & \ & \lambda_{22} & \ & & \\
  & \ & \ & \  & \lambda_{11} & \ & \ & &
\end{array}
\label{GTintroduced}
\end{equation}
subject to the constraints $\lambda_{ki}\geq \lambda_{k-1,i}\geq \lambda_{k,i+1}$. Note that $\lambda_{Nk}=\lambda_k$. The array $\Lambda$ is called a GT pattern. Given that the $\gla(1)$ representations are of dimension one and that the decomposition of a $\gl_k$ representation into $\gl_{k-1}$ is multiplicity free, it is clear that one can form  a basis of $\lV_\lambda$ parametrised by all the GT patterns. The pattern where all nodes take the maximal/minimal possible value corresponds to the highest-/lowest-weight state of $\lV_{\lambda}$.

To understand which operators of $U(\gla(N))$ are diagonal in the GT basis, define $\mathbb{G}_N$ through the following column-ordered determinant
\begin{equation}
\label{defG}
\mathbb{G}_N(u)=\det_{1\leq i,j \leq N}\left[\left(u\,\delta_{ij}-\hbar E_{ji}\right)e^{-\hbar\partial_u}\right]e^{\hbar N\partial_u}\,.
\end{equation}
The coefficients of $\mathbb{G}_N(u)$ expanded in powers of $u$ are the Casimir operators generating  the center of $U(\gla(N))$. Their values uniquely define the highest-weight representation. Instead of expanding in $u$, let us factorise:
\begin{equation}
\mathbb{G}_N(u)=\prod_{i=1}^{N}(u-\hbar\,(\hat\lambda_{Ni}+N-1))\,,
\end{equation}
where the constant shift $N-1$ was introduced for further convenience.

Strictly speaking, $\hat\lambda_{Ni}$ are not elements of $U(\gla(N))$, only their symmetric combinations are - this is the Harish-Chandra isomorphism \cite{Harish}. However, it is not a problem as for every particular irreducible representation they become number-valued operators, so we ignore this subtlety.

The value of operators $\hat\lambda_{Ni}$ is the same on the whole representation space, so we can find it by action on some particular vector. It is convenient to choose the lowest-weight vector since $E_{ij}|{\rm LWS}\rangle=0$ for $i>j$ and then the determinant is column-ordered \eqref{defG},  it becomes computable by the product of the matrix's diagonal entries. Provided that the weight of the lowest-weight state is $[\lambda_N,\lambda_{N-1},\ldots,\lambda_1]$, one finds
\begin{equation}
\mathbb{G}_N(u)=\prod_{i=1}^N(u-\hbar(\lambda_{i}+N-i))\,,
\end{equation}
so that $\hat\lambda_{Ni}=\lambda_i+1-i\,.$ These are the so-called shifted weights.

Similarly define $\mathbb{G}_k(u)=\det_{1\leq i,j \leq k}\left[\left(u\,\delta_{ij}-\hbar E_{j,i}\right)e^{-\hbar\partial_u}\right]e^{\hbar k\partial_u}$ and compute $\hat\lambda_{ki}=\lambda_{ki}+1-i$ for $1\leq k\leq N-1$. As $\mathbb{G}_k(u)$ generates the center inside $U(\gla(k))\subset U(\gla(N))$, it is clear that $\mathbb{G}_k$ for $1\leq k\leq N$ generate a commutative subalgebra inside $U(\gla(N))$ known as the GT algebra. The GT algebra acts diagonally on the GT basis:
\begin{equation}
\mathbb{G}_k(u)|\Lambda\rangle =\prod_{i=1}^k (u-\hbar(\hat\lambda_{ki}+k-1))|\Lambda\rangle\,.
\end{equation}
We note that Cartan subalgebra of $\gla(N)$ is a part of the GT algebra. Its action on the GT basis follows from
\begin{equation}
\sum_{i=1}^kE_{ii}\ket{\Lambda}=\sum_{i=1}^k \lambda_{ki}\ket{\Lambda}\,,
\end{equation}
and it is clear that its spectrum is degenerate. For example, consider the $[2,1,0]$ irrep of $\gla(3)$ used to describe the baryon octet in particle physics. Then the patterns 
\begin{equation}
\begin{array}{ccccc}
2 & & 1 & & 0 \\
 & 2 & & 0 & \\
 & & 1 & & 
\end{array}, \quad 
\begin{array}{ccccc}
2 & & 1 & & 0 \\
 & 1 & & 1 & \\
 & & 1 & & 
\end{array}
\end{equation}
both correspond to basis vectors with weight $[1,1,1]$ and hence are indistinguishable from the point of the of view of the Cartan subalgebra. But they are distinguishable using the GT algebra.

This goes in contrast to the $\gla(2)$ case where the eigenvalue of $E_{11}-E_{22}$ uniquely determines a vector in an irrep, and this is a major reason for why generalisations of $\gl(2)$ results to higher ranks are not always possible or obvious. The only representations where the Cartan subalgebra of $\gl(N)$ has non-degenerate spectrum are the symmetric $(S^1)$ and anti-symmetric $(1^A)$ powers of the defining representation and their conjugates $(S^{N-1})$, $(1^{N-A})$. As we shall see, it is easier to make claims for these representations, and it is unlikely to be a coincidence.

\begin{figure}
\centering
  \includegraphics[width=60mm,scale=0.5]{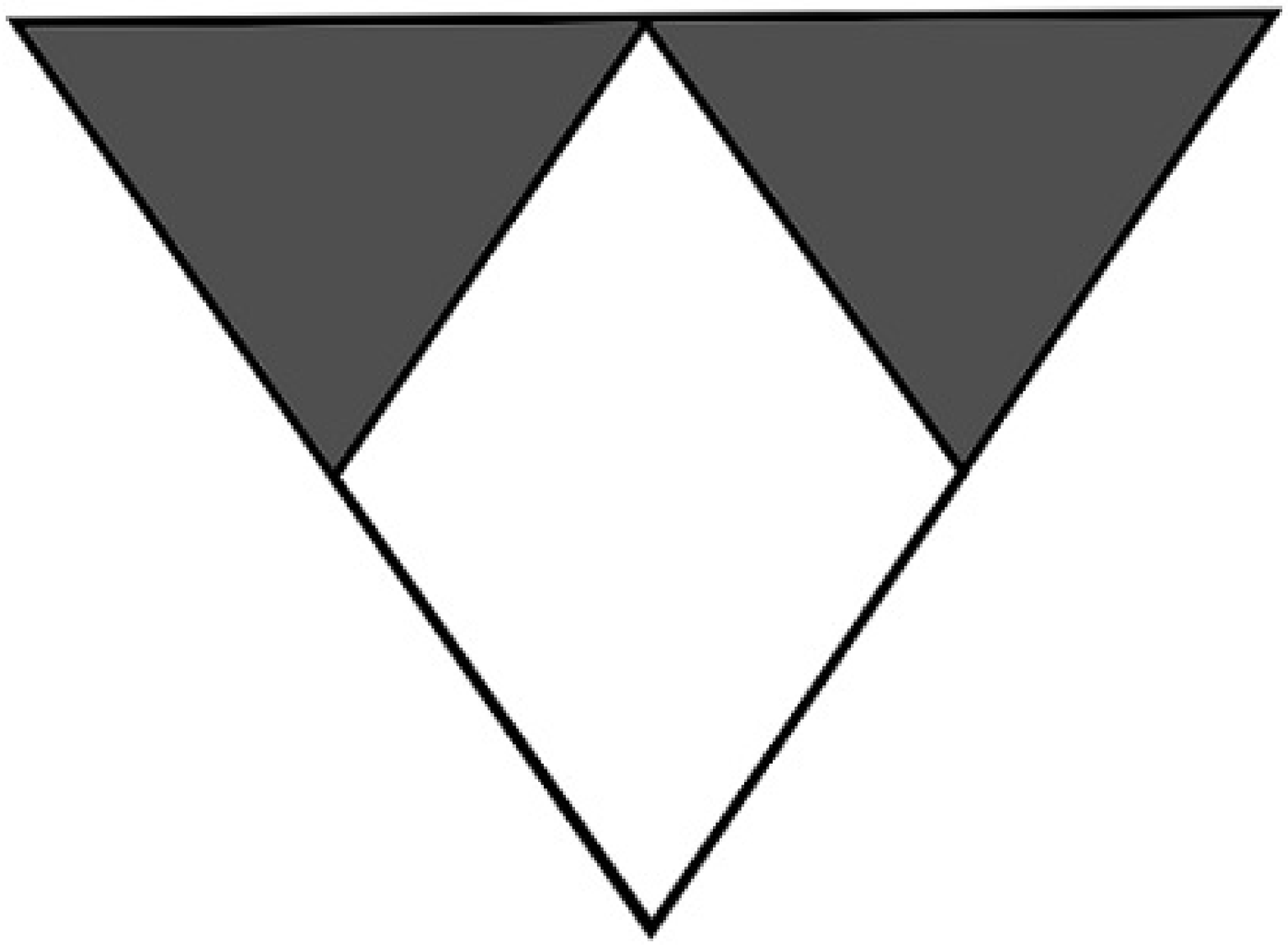}
  \caption{Free (white) diamond-shaped area on a GT pattern. For the $(S^A)$ representation, the nodes in the left shaded area are fixed to $S$ and the nodes in the right shaded area are fixed to $0$. }
  \label{fig:diamond2}
\end{figure}

For a large section of this work we will be interested in \textit{rectangular} representations $(S^A)$. For these representations the branching rules fix a large section of the GT patterns with only a free diamond-shaped area in the middle, see Figure \ref{fig:diamond2}.
It is then convenient to map this diamond to a rectangle 
\begin{equation}
\label{mintro}
\raisebox{1.5em}{
$
\begin{array}{cccc}
\lambda_{N-A,1} & \lambda_{N-A+1,2} & \dots & \lambda_{N-1,A} \\
\lambda_{N-A-1,1} & \lambda_{N-A,2} & \dots & \lambda_{N-2,A} \\
\vdots & \vdots & \ddots & \vdots \\
\lambda_{1,1} & \lambda_{2,2} & \dots & \lambda_{A,A} 
\end{array}
\equiv
\begin{array}{cccc}
m_{N-A,1} & m_{N-A,2} & \dots & m_{N-A,A} \\
m_{N-A-1,1} & m_{N-A-1,2} & \dots & m_{N-A-1,A} \\
\vdots & \vdots & \ddots & \vdots \\
m_{1,1} & m_{1,2} & \dots & m_{1,A} 
\end{array}
$
}\,,
\end{equation}
where on the right we have introduced a relabelling $m_{kj}=\lambda_{k+j-1,j}$ of the nodes which will be useful later.

\paragraph{GT basis for $Y(\gla(N))$} 
A length $L$ spin chain that we consider is just a product of $L$ copies of the $\gla(N)$ irrep $\lV$, hence we can in principle just use the $L$-tuple of GT patterns, with each pattern addressing one spin chain site. However, while the labelling of a Hilbert space basis using these patterns turns out to be the right one, the associated eigenvectors do not relate properly to the Yangian generators so it is unlikely they are the eigenvectors of good separated variables. In $N=2$ such an idea would correspond to erasing the ${\rm Nil_L}$ part in \eqref{eq:T11spec} which clearly makes $\B$ unrelated to the Bethe algebra. Hence we will be more accurate and use the construction appropriate for the Yangian representations.

Let us note that the Yangian $Y(\gla(N))$ represented on an $L=1$ spin chain coincides with the corresponding representation of $U(\gla(N))$ and the operator $\mathbb{G}_N$ defined by \eqref{defG} is precisely the quantum determinant which is equal to $T[^{12\dots N}_{12\dots N}](u)$, {\it cf.}  \eqref{YangCappeli} with $1$ being dropped out in the determinant. The quantum determinant is the generator of the Yangian's center hence it is a correct object for generalisation for longer spin chains. 

Consider a sequence of imbeddings $Y(\gla(1))\subset Y(\gla(2))\subset\dots\subset Y(\gla(N))$ given by injection $T_{ij}(u)\rightarrow T_{ij}(u)$. The centre of the $Y(\gla(k))$ term is generated by the corresponding quantum minor $T[^{12\dots k}_{12\dots k}](u)$. The GT algebra \cite{Molev1994} is then defined as the commutative algebra generated by all quantum minors $T\left[^{12\dots k}_{12\dots k}\right](u)$, $k=1,\dots,N$. As we discussed, it coincides with the singular twist Bethe algebra $\mathcal{B}^H$. 

The GT basis is the basis in which all elements of the GT algebra act diagonally. Denote an element in this basis by
\begin{equation}
\ket{\Lambda_1,\dots,\Lambda_L}^{\rm GT}\,,
\end{equation}
where $\Lambda_\alpha$ is a GT pattern with the nodes $\lambda_{nm}^{\alpha}$, and $\lambda_{Nm}^{\alpha}=\nu_m$, $m=1,\dots,N$. If no ambiguity is possible, a short-hand notation $\ket{\Lambda}^{\rm GT}$ will still be used.

Define polynomials 
\begin{equation}
\lambda_{ij}(u)=\prod_{\alpha=1}^L(u-\theta_\alpha-\hbar\lambda_{ij}^\alpha)\,.
\end{equation}
Then 
\begin{equation}
T^{[2a-2]}\left[^{1\dots a}_{1\dots a}\right](u)\ket{\Lambda_1,\dots,\Lambda_L}^{\rm GT}=\prod_{k=1}^a \lambda_{ak}^{[2k-2]}(u)\ket{\Lambda_1,\dots,\Lambda_L}^{\rm GT}\,.
\end{equation}
This then implies that the value of any node $\lambda_{ak}^\alpha$ appearing in the tuples $\Lambda_1,\dots,\Lambda_L$ can be read off from the eigenvalue of $T\left[^{1\dots a}_{1\dots a}\right](u)$. Note that the basis $\ket{\Lambda_1,\dots,\Lambda_L}$ is naturally factorised -- each $\Lambda_\alpha$ is independent of $\Lambda_\beta$ for $\alpha\neq \beta$ and, although we will not need them in the present work, raising and lowering operators satisfying (\ref{eq:Px}) can also be constructed in a canonical way. Due to these properties, the GT algebra is extremely plausible candidate for a separated variable algebra of $\gl(N)$ spin chains. And indeed it was considered in this perspective in \cite{Valinevich:2016cwq}, including a more general class of Yangian representations than we address here. However, factorisation of wave functions for physically relevant eigenvectors of the Bethe algebra $\mathcal{B}^G$ was not shown. In fact it does not happen in the GT basis, save for $N=2$ case, as we are going to discuss.

\subsection{Spectrum of $\B(u)$\label{sec:SpectrumOfB}}
For $N=2$ it was possible to find the reference frame where $\B$ is identical with $\wT^H$ and hence is a member of the GT algebra. Now we attempt to generalise this property to higher rank.  $\B(u)$ is a polynomial of degree $L\frac{N(N-1)}{2}$, which is the sum of polynomial degrees of $\wT_{a,1}^H(u)$ with $1\leq a\leq N-1$, so we expect that $\B(u)$ is a product of these singular twist transfer matrices. Note that we have excluded $\wT_{N,1}^H\propto T[^{12\dots N}_{12\dots N}](u)$ from consideration since it is just a number-valued function on the Yangian irrep and hence cannot distinguish eigenstates. For the case when $\lV_\lambda$ is the defining representation, the spectrum of $\B(u)$ was propsed in \cite{Gromov:2016itr}. After appropriate adjustments in conventions we conclude that this spectrum is identical with the spectrum of $\BGT$ defined by
\begin{equation}
\label{BGT}
\BGT \propto \displaystyle T\left[^{1}_{1}\right]T^{[2]}\left[^{1\ 2}_{1\ 2}\right]\dots T^{[2N-4]}\left[^{1\ 2\ \dots N-1}_{1\ 2\ \ldots N-1}\right]\,,
\end{equation}
and normalised to be a monic polynomial in $u$.

Hence we question  whether it is possible to find such $G$ that $\B^{G}=\BGT$. By considering $L=1$ and the defining representation, $G$ is already almost uniquely fixed to be the companion matrix \eqref{eq:Gcompanion} by our request, save for some inessential residual freedom in $G$. However for either $L>1$ or representations more involved than the defining one, it is no longer true that $\B^{G}=\BGT$, except for $N=2$. However, while the answer to our question is negative, this still proved to be a useful exercise because the companion twist frame has numerous remarkable properties. We will now begin investigating them.

First, let us compute $\B$ explicitly in this frame. In component form our twist is given by
\begin{equation}
G_{kj}=(-1)^{j-1}\chi_j\delta_{k1}+\delta_{k,j+1}\,,
\label{twsitG}
\end{equation}
while the dressed minors in \eqref{Bgooddef} are related to the bare ones by 
\begin{equation}
\label{eq:some1}
{\bf T}\left[^{j_n}_{j_{n-1}N}\right]=\displaystyle\sum_{k}T\left[^{j_n}_{k} \right]G_{k_1 j^1_{n-1}}\dots G_{k_{n-1} j^{n-1}_{n-1}}G_{k_n N}\,,
\end{equation}
where $k=k_1k_2\ldots k_n$ and $j_n=j_{n}^1\ldots j_{n}^n$ are multi-indices.
Since $G_{k_n N}=(-1)^{N-1}\chi_N\delta_{k_n1}$, the only terms which can survive in \eqref{eq:some1} are those with $k_n=1$. But then, by the antisymmetry of the quantum minor, we have that only $k_i>1$ terms survive, $i=1,\dots,n-1$, in which case $G_{k_i,l}=\delta_{k_i,l+1}$, and we  find 
\begin{equation}
{\bf T}\left[^{j_n}_{j_{n-1}N}\right]= (-1)^{n+N-2}\chi_N\, T\left[^{j_n}_{1\ j_{n-1}+1}\right]\,,
\end{equation}
where $j_{i}+1\equiv j_{i}^1+1,j_{i}^2+1,\ldots,j_{i}^i+1$.

Hence
\begin{equation}
\label{BCgood}
\fbox{
\raisebox{0.2em}{
$\B(u)\propto \displaystyle\sum_{k}T\left[^{k_1}_{1}\right]T^{[2]}\left[^{k_2}_{1\ k_1+1}\right]\dots T^{[2N-4]}\left[^{1\ 2\ \dots N-1}_{1\ k_{N-2}+1}\right]
$
}\,.
}
\end{equation}

Explicit expressions are $\B= T[^1_1]$ for $N=2$ and $\B\propto T[^1_1]T^{[2]}[^{12}_{12}]+T[^2_1]T^{[2]}[^{12}_{13}]$ for $N=3$.

Now we claim that
\begin{eqnarray}
\B=\BGT+{\rm Nil_C}\,,
\end{eqnarray}
where by ${\rm Nil_C}$ we mean an operator which becomes an upper-triangular matrix in a properly ordered GT basis.

\noindent{\it Proof:} It suffices to present a partial order $\succ$ in the GT basis such that action of each summand $\CO=T\left[^{k_1}_{1}\right]T^{[2]}\left[^{k_2}_{1\ k_1+1}\right]\dots T^{[2N-4]}\left[^{1\ 2\ \dots N-1}_{1\ k_{N-2}+1}\right]$ in \eqref{BCgood}, save for the $\BGT$ term, is strictly positive, or zero. That is if $\CO|\Lambda\rangle^{\rm GT}=\sum_{r} c_r |\Lambda_r\rangle^{\rm GT}$, $c_r\neq 0$ then $|\Lambda_r\rangle^{\rm GT}\succ|\Lambda\rangle^{\rm GT}\,,\ \forall r$.

We will partially order the GT basis vectors with respect to their Cartan weight $\lambda=[\lambda_1,\ldots,\lambda_N]$, i.e. $\pi^\CH(E_{ii})|\Lambda\rangle^{\rm GT}=\lambda_i|\Lambda\rangle^{\rm GT}$, where $\pi^\CH(E_{ii})\equiv \sum_{\alpha=1}^L\pi^{\nu}_\alpha(E_{ii})$. One defines $\lambda\succ\lambda'$ iff $\lambda_i>\lambda_i'$ for the smallest $i$ for which $\lambda_i\neq \lambda_i'$. 

Define $A=[^\alpha_\beta]$, where $\alpha=\{k_1,k_2,\ldots,k_{N-2},1,2,\ldots, N-1\}$ and $\beta=\{1,1,k_1+1,1,k_2+1,\ldots,k_{N-2}+1\}$. Then 
\begin{eqnarray}
\label{ECO}
[\pi^{\CH}(E_{ii}),\CO]=\left(\sum_{b\in\beta}\delta_{i,b}-\sum_{a\in\alpha}\delta_{i,a}\right)\CO\,.
\end{eqnarray}
Construct $A^{\rm reg}=[^{\alpha_{\rm reg}}_{\beta_{\rm reg}}]$  from $A$ by repeated crossing out of pairs $(\alpha_i,\beta_j)$, where $\alpha_i=\beta_j$, until no such pairs left. Replacing $A$ with $A^{\rm reg}$ won't affect \eqref{ECO}. 

The only possibility that $A^{\rm reg}=[^{\emptyset}_{\emptyset}]$ is  $\CO=\B^{\rm GT}$. For all other cases it is easy to see that $\min[\beta_{\rm reg}]<\min[\alpha_{\rm reg}]$ which clearly implies that action of $\CO$ on the elements of the GT basis, if non-zero, is strictly positive in the above-defined sense. \qed\\[0em]

Now we are ready to introduce the operators $X_{\svbas}$  which will turn out to be the good separated variables. One can follow the following logic: Whereas $\B$ and $\BGT$ do not coincide, their spectrum is nevertheless equal, since ${\rm Nil_C}$ is upper-triangular. Therefore we will label the eigenvectors of $\B$ by $|\Lambda\rangle$, where $\Lambda$ is a collection of $L$ GT patterns for representation $\lV$, but note that $|\Lambda\rangle\neq |\Lambda\rangle^{\rm GT}$ in general. We introduce labelling of the operatorial zeros of $\B$ by
\begin{equation}
\label{eq:BX}
\B=\prod_{\alpha=1}^L\prod_{k=1}^{N-1}\prod_{j=1}^{N-k}(u-X_{kj}^\alpha)\,,
\end{equation}
with their spectrum given by
\begin{equation}
\label{Xegvalues}
\fbox{
$\displaystyle
X_{kj}^\alpha|\Lambda_1,\Lambda_2,\ldots,\Lambda_L\rangle = \left(\theta_\alpha+\hbar\,\hat\lambda_{k+j-1,j}^{\alpha}\right)|\Lambda_1,\Lambda_2,\ldots,\Lambda_L\rangle\,,
$
}
\end{equation}
where $\hat\lambda_{ij}^{\alpha}=\lambda_{ij}^{\alpha}-j+1\,,$ and $\lambda_{ij}^{\alpha}$ is an element of the $\alpha$'s GT pattern \eqref{GTintroduced}, with $\lambda_{Ni}^{\alpha}=\nu_i$.\\[0em]

The presented logic has however some subtleties and weak points. By subtlety we mean a statement that requires further clarification and by weak point we mean a statement that requires further arguments to prove being correct.

One technical subtlety is that $\B$ is a symmetric polynomial in $X$'s, so in \eqref{Xegvalues} we actually agree on a way to define {\it e.g.} $X_1,X_2$ from their known combinations $X_1+X_2,X_1X_2\,.$

The second subtlety is that $\BGT$ has a degenerate spectrum, even assuming that $\theta_{\alpha}$ are distinct. Indeed, $\BGT$ is only a product of operators generating the GT algebra, so it bears less information. The special cases when $\B^{GT}$ is still non-degenerate are the rectangular representations with $A=1$, $A=N-1$, or $S=1$, that is same ones when the Cartan subalgebra is non-degenerate. Similarly to $\BGT$, $\B$  turns out to be degenerate as well, and hence not all $\ket{\Lambda}$ are uniquely defined from the fact that they are $\B$-eigenvectors. One should provide a separate prescription which basis diagonalising $\B$ we would like to choose.

The first weak point is that it was never proven that $[\B(u),\B(u')]=0$ for arbitrary $u,u'$. In fact, there is already an example of a super-symmetric analog of $\B$ \cite{Gromov:2018cvh} which is non-commuting. In our case, it turns out that $\B(u)$ and $\B(u')$ commute indeed, but we  shall not prove commutativity of $\B$ directly. Instead we demonstrate it later by constructing the basis of $\ket{\Lambda}$'s in a $u$-independent way.

The second weak point is that $\B$ was not proven to be diagonalisable, and in principle it might be not the case as $\B$ is equal to the degenerate diagonal matrix $\BGT$ added with an upper-triangular matrix. Again, we shall not prove diagonalisability of $\B$ directly, but this will follow after we construct enough of $\ket{\Lambda}$'s that will turn out to be the eigenvectors of $\B$.

We therefore see that it is not enough to consider \eqref{Xegvalues} simply as a consequence of \eqref{eq:BX}. For the above-outlined reasons, we need to construct the basis of $\ket{\Lambda}$'s in an independent way. We offer such a construction in the next section. Furthermore we will get the results about factorisation of the wave functions that allow one to indeed consider  $X$'s as separated variables.

\section{Separation of Variables}
Let us make a remark on the notation to be used. In the following we will actively use both bra-vectors $\langle v|$ and ket-vectors $|w\rangle$. Ket-vectors denote elements of the Hilbert space $\mathcal{H}$, and we will construct eigenvectors of the Bethe algebra $\mathcal{B}^G$ as ket-vectors. Bra-vectors are elements of the dual space $\mathcal{H}^*$, and we will construct eigenvectors of the separated variables $X_{ij}^\alpha$ as bra-vectors. The notation $\langle v|w\rangle$ means the canonical pairing between a vector and a dual vector. If $|\Lambda\rangle$ is an element of a basis then we define $\langle \Lambda|$ as an element of the dual basis, i.e. by $\langle \Lambda|\Lambda'\rangle =\delta_{\Lambda\Lambda'}$. Obviously, if an operator $X$ is diagonal in this basis then $X|\Lambda\rangle=\theta|\Lambda\rangle$ implies $\langle \Lambda|X=\langle \Lambda|\theta$, so formulae like \eqref{Xegvalues} won't change when switching to the dual space. 

For operators represented by non-diagonal matrices in a basis, one should take the transpose of the matrix when switching to the dual basis. Notice in particular that ${\rm Nil_C}|{\rm HWS}\rangle^{\rm GT}=0$ but $\bra{\rm LWS}^{\rm GT}{\rm Nil_C}=0\,,$ where $\ket{\rm HWS}^{\rm GT}/\ket{\rm LWS}^{\rm GT}$ are, respectively, the highest-/lowest- weight states of the Yangian representation.

\subsection{Construction of SoV basis for $N=2$}
In this subsection we will outline a procedure for constructing an SoV basis for $Y(\gla(2))$, in a way that reproduces the proposal of {\cite{Maillet:2018bim}} for the defining representation and  also makes it precise for higher-spin representations. This will serve as a precursor for our study of the higher-rank cases. We omit some of the proofs as they are either available in the literature or follow  naturally as specialisations of the forthcoming more general discussion.

For the companion twist, we have 
\begin{equation}
\B=T_{11}(u),\quad \T_{1,1}(u)=\chi_1 T_{11}(u)+T_{12}(u)-\chi_2 T_{21}(u)\,.
\end{equation}
We introduce $\bra{0}\equiv \bra{\rm LWS}^{\rm GT}$ -- the dual of the lowest-weight state of the Yangian irrep. It corresponds to the $L$-tuple of GT patterns with all  $\lambda_{11}^{\alpha}=\nu_2$, and we set $\nu_2=0$ for simplicity. This state satisfies 
\begin{equation}
\bra{0}T_{j1}(u)=\delta_{j1}Q_\theta(u)\bra{0}\,.
\end{equation}
We then obtain that the action of $\T_{1,1}(\theta_\alpha)$ on $\bra{0}$ simplifies to 
\begin{equation}
\bra{0}\T_{1,1}(\theta_\alpha)=\bra{0}T_{12}(\theta_\alpha)\,.
\end{equation}
By using RTT, one shows that $\bra{0}$ is an eigenvector of $\B$ with the eigenvalue 
\begin{equation}
(u-\theta_\alpha-\hbar)\prod_{\beta\neq\alpha}(u-\theta_\beta)\,.
\end{equation}
As a result, this state is annihilated by the subsequent action of $T_{11}(\theta_\beta)$, $\beta\neq\alpha$. 
Also this state is annihilated by $T_{21}(\theta_\beta)$, which can be easily checked using RTT. Hence
\begin{equation}
\label{indepinhom}
\bra{0}\T_{1,1}(\theta_\alpha)\T_{1,1}(\theta_\beta)=\bra{0}T_{12}(\theta_\alpha)T_{12}(\theta_\beta)\,,\quad \alpha\neq \beta\,
\end{equation}
which can be shown to be an eigenvector of $\B$ with the eigenvalue $
(u-\theta_\alpha-\hbar)(u-\theta_\beta-\hbar)\prod_{\gamma\neq \alpha,\beta}(u-\theta_\gamma)\,.$
Then, by induction, it follows that for any subset $I\subset \{1,2,\dots,L\}$ we have that 
\begin{eqnarray}
&&\bra{0}\prod_{\alpha\in I}\T_{1,1}(\theta_\alpha)=\bra{0}\prod_{\alpha\in I}T_{12}(\theta_\alpha)\,,\quad {\rm and}\\ \nonumber
&&\bra{0}\prod_{\alpha\in I}\T_{1,1}(\theta_\alpha)\,\B(u)=\prod_{\alpha\in I}(u-\theta_\alpha-\hbar)\prod_{\beta\notin I}(u-\theta_\beta)\bra{0}\prod_{\alpha\in I}\wT_{1,1}(\theta_\alpha)\,.
\end{eqnarray}
Furthermore, all of these states are non-zero, since $\T_{1,1}(u)$ has no vanishing eigenvalues at $u=\theta_\alpha$. In this manner we can construct $2^L$ states. This precisely matches the dimension of the Hilbert space if $\lV$ is the defining representation. Hence, for the case of the defining representation, the constructed states form a basis, since each corresponds to a different eigenvalue of $\B$. 

An important point very useful for generalisations is that the constructed states are independent of the twist eigenvalues, as they should be since $\B$ is independent of these and so naturally its eigenvectors are as well. We will routinely make use of this fact.

For a more general case of symmetric power representation $\nu=[S,0]$, the constructed states are not sufficient to span the Hilbert space, and we should look for more. 

As noted in \cite{Maillet:2018bim}, it is natural to conjecture that the basis is not constructed just with $\T_{1,1}(\theta)$, but also with $\T_{1,1}(\theta+n \hbar)$, $n\in\ZZ$. Indeed, this is analogous to the way the GT basis is constructed \cite{molev2007yangians} -- a generic eigenvector of $T_{11}(u)$  can be obtained by acting on $\bra{0}$ with $T_{12}(\theta)T_{12}(\theta+\hbar)T_{12}(\theta+2\hbar)\dots$. To put it more in the perspective of a physicist, one can introduce operators $X_{\alpha}$ as operatorial zeros of $T_{11}(u)$ whose spectrum was described after \eqref{eq:T11spec}. One then finds, using RTT, that  the ladder operators  are
\begin{equation}
\label{ladderoperators}
P^+_\alpha= T_{12}(X_{\alpha})\,,\quad P^-_{\alpha}= T_{21}(X_{\alpha})\,,
\end{equation}
so that 
\begin{equation}
\bra{s_{\alpha}}\equiv \bra{0}(P^+_{\alpha})^s =\bra{0}T_{12}(\theta_{\alpha}+\hbar)\ldots T_{12}(\theta_{\alpha}+\hbar\,(s-1)\hbar)\,
\label{ladderrep}
\end{equation}
is the $\B$-eigenstate. Normal ordering is used in the above expressions, that is $X$'s are placed to the left of other operators.

Still, the representation \eqref{ladderrep} of the $\B$-eigenstates is not fully satisfactory as it does not suggest yet that the wave functions would factorise in this basis, so we would like to replace $T_{12}$ with transfer matrices in order to conclude about factorisation.

The action of $\T_{1,1}(\theta_\alpha)$ (once) is equivalent to the action of $P^+_{\alpha}$ (once) on $\bra{0}$, as we learned above. However, the action of $\T_{1,1}(\theta_\alpha)\T_{1,1}(\theta_\alpha+\hbar)$ on $\bra{0}$ does not yield $\bra{0}(P_\alpha^+)^2$ as we would like. Instead, it yields
\begin{equation}
\bra{0}\T_{1,1}(\theta_\alpha)\T_{1,1}(\theta_\alpha+\hbar)=\bra{0}T_{12}(\theta_\alpha)T_{12}(\theta_\alpha+\hbar)-\chi_2\bra{0}T_{12}(\theta_\alpha)T_{21}(\theta_\alpha+\hbar)\,.
\end{equation}
The second term is non-vanishing, as can be checked using RTT. Instead, it can be rewritten in a useful form 
\begin{equation}
-\chi_2\bra{0}\T_{12}(\theta_\alpha)T_{21}(\theta_\alpha+\hbar)=\chi_2 \bra{0}\left(T_{11}(\theta)T_{22}(\theta_\alpha+\hbar)-T_{12}(\theta_\alpha)T_{21}(\theta_\alpha+\hbar) \right)
\end{equation}
since the first term on the r.h.s. vanishes. We can recognise the transfer matrix $\T_{2,1}$ in the expression on the r.h.s., and so we can see that the eigenstate $\bra{2_\alpha}$ is actually given by 
\begin{equation}
\bra{2_\alpha}=\bra{0}(\T_{1,1}(\theta_\alpha)\T_{1,1}(\theta_\alpha+\hbar)-\T_{2,1}(\theta_\alpha+\hbar))=\bra{0}\T_{1,2}(\theta_\alpha)\,,
\end{equation}
where  the Hirota equation  \cite{Zabrodin:1996vm} enjoyed by the transfer matrices was used on the last step.

The $\B$-eigenvalue of $\bra{2_{\alpha}}$ is $(u-\theta_\alpha-2\hbar)\prod_{\beta\neq \alpha}(u-\theta_\beta)\,.$ It is then natural to guess that, in order to construct the eigenstate $\bra{s_{\alpha}}$ of $\B$ with eigenvalue $(u-\theta_\alpha-s\ \hbar)\prod_{\beta\neq \alpha}(u-\theta_\beta)$,
we should act on $\bra{0}$ with $\T_{1,s}(\theta_\alpha)$, and therefore
\begin{equation}
\label{T12T11}
\bra{0}\wT_{1,s}(\theta_\alpha)=\bra{0}T_{12}(\theta_\alpha)\dots T_{12}(\theta_\alpha+\hbar(s-1))\,.
\end{equation} 
This can be  seen for instance by recursively using relation \cite{Zabrodin:1996vm}
\begin{equation}
\label{Zrec}
\T_{1,s+1}(u)=\T_{1,s}(u)\T_{1,1}(u+\hbar\ s)-\T_{1,s-1}(u)\T_{2,1}(u+\hbar\ s)\,.
\end{equation}
We won't present this computation here, but instead give a quick argument supporting \eqref{T12T11}. As the result is not expected to depend on twist, let us set $\chi_1=\chi_2=0$ which sets the second term in \eqref{Zrec} to zero and also simplifies $\wT_{1,1}(u)$ to  $\wT_{1,1}(u)=T_{12}(u)$. Then the desired property \eqref{T12T11} is demonstrated immediately by recursion. 

We can also produce formulae of type \eqref{indepinhom} and finally conclude that all eigenstates of $\B$, which we label by $\bra{\Lambda}=\bra{s_1,\ldots, s_L}$ for $s_\alpha\in \{0,1,\dots,S\}$, can be constructed as
\begin{equation}
\label{eq:Lambdadir}
\bra{\Lambda}=\bra{0}\prod_{\alpha=1}^L \T_{1,s_\alpha}(\theta_\alpha)\,.
\end{equation}
Their $\B$-eigenvalues are
\begin{equation}
\bra{\Lambda}\B(u)=\prod_{\alpha=1}^L (u-\theta_{\alpha}-\hbar\,s_\alpha)\bra{\Lambda}\,.
\label{eq:egvaluesf}
\end{equation}
In the case of $N=2$, we were quite lucky to know the ladder operators \eqref{ladderoperators}, so deriving \eqref{T12T11} was sufficient for demonstration of \eqref{eq:egvaluesf}. For higher-rank cases, we won't be able to get a straightforward generalisation of \eqref{ladderoperators}. Instead we will develop a related but more flexible approach to show that the generalisation of \eqref{eq:Lambdadir} are eigenvectors of $\wB$.

\subsection{Commutation relation between $\B$ and $\wT_{\lambda}$}
From the results obtained for the $\gl(2)$ case, we can expect that eigenstates of $\wB$ are generated by some action of transfer matrices on the state $\bra{0}\equiv \bra{\rm LWS}^{\rm GT}$ which we call the GT vacuum. It is natural to assume that we should not restrict ourselves only to symmetric representations when considering higher rank $\gl(N)$ symmetry, so we will study transfer matrices in arbitrary finite-dimensional representations. These reprsentations are parameterised by integer partitions of length at most $N$, $\lambda=[\lambda_1,\lambda_2,\ldots,\lambda_N]$ or, equivalently, by Young diagrams.

Since $\B$ does not depend on $z_1,\ldots, z_N$, its eigenvectors cannot depend on $z$'s either. Hence we start our investigation by considering the null twist $G=\mathcal{N}$, that is we put all $z_i=0$. This convention is assumed until the end of this subsection. In the next subsection, we will discuss the mechanism by which the twist dependence indeed cancels out.

Transfer matrices play a role somewhat similar to the ladder operators for $\B$, {\it cf.}~\eqref{T12T11}. Hence it is instructive to find their commutation relations with $\B$. To this end we shall be using graphical notations.

To get technical simplifications at later stages, in particular to avoid needing to invert fused R-matrices, we  use, up to a scalar factor, the inverse $R$-matrix to perform scattering between two defining representations:
\begin{eqnarray}
\raisebox{-0.4\height}{
\begin{picture}(40,20)(0,0)
\put(0,5){$v$}
\put(0,21){$u$}
\put(0,0){\vector(2,1){40}}
\put(0,20){\vector(2,-1){40}}
\end{picture}
}
\equiv (u-v)\Id+\hbar \Perm\,.
\end{eqnarray}
Then, if we denote the bare monodromy matrix defined by \eqref{eq:defT}  as
\begin{eqnarray}
\raisebox{-0.4\height}{
\begin{picture}(40,20)(0,-10)
\put(0,0){\line(1,0){13}}
\put(27,0){\vector(1,0){13}}
\polygon(13,4)(27,4)(27,-4)(13,-4)
\put(17,-2){$u$}
\put(3,5){$i$}
\put(33,5){$j$}
\end{picture}
}
\equiv T_{ij}(u)\equiv T[^i_j](u)\,,\quad \text{equivalently}\,\quad
\raisebox{-0.4\height}{
\begin{picture}(40,20)(0,-10)
\put(0,0){\line(1,0){13}}
\put(27,0){\vector(1,0){13}}
\polygon(13,4)(27,4)(27,-4)(13,-4)
\put(17,-2){$u$}
\end{picture}
}
=
\sum_{ij}E_{ij}\otimes T_{ij},
\end{eqnarray}
the $TTR^{-1}=R^{-1}TT$ relation would read
\begin{eqnarray}
\raisebox{-0.4\height}{
\begin{picture}(60,20)(0,-10)
\put(00,-5){
\put(0,0){\line(1,0){13}}
\put(27,0){\line(1,0){3}}
\polygon(13,4)(27,4)(27,-4)(13,-4)
\put(17,-2){$v$}
}
\put(00,5){
\put(0,0){\line(1,0){13}}
\put(27,0){\line(1,0){3}}
\polygon(13,4)(27,4)(27,-4)(13,-4)
\put(17,-2){$u$}
}
\put(30,0){
\qbezier(0,-5)(5,-5)(15,0)
\qbezier(15,0)(25,5)(30,5)
\qbezier(0,5)(5,5)(15,0)
\qbezier(15,0)(25,-5)(30,-5)
}
\end{picture}
}
=
\raisebox{-0.4\height}{
\begin{picture}(60,20)(0,-10)
\put(30,-5){
\put(0,0){\line(1,0){3}}
\put(17,0){\line(1,0){13}}
\put(-10,0){
\polygon(13,4)(27,4)(27,-4)(13,-4)
}
\put(7,-2){$u$}
}
\put(30,5){
\put(0,0){\line(1,0){3}}
\put(17,0){\line(1,0){13}}
\put(-10,0){
\polygon(13,4)(27,4)(27,-4)(13,-4)
}
\put(7,-2){$v$}
}
\qbezier(0,-5)(5,-5)(15,0)
\qbezier(15,0)(25,5)(30,5)
\qbezier(0,5)(5,5)(15,0)
\qbezier(15,0)(25,-5)(30,-5)
\end{picture}
}
\,.
\label{RTTgraph}
\end{eqnarray}

For contraction of indices, we always implicitly assume insertion of the null twist, and hence the following relation holds:
\begin{eqnarray}
\raisebox{-0.4\height}{
\begin{picture}(70,20)(0,-10)
\put(0,0){\line(1,0){13}}
\put(27,0){\line(1,0){3}}
\polygon(13,4)(27,4)(27,-4)(13,-4)
\put(17,-2){$u$}
\put(30,0){
\put(0,0){\line(1,0){13}}
\put(27,0){\vector(1,0){13}}
\polygon(13,4)(27,4)(27,-4)(13,-4)
\put(17,-2){$u$}
}
\put(3,5){$i$}
\put(63,5){$j$}
\end{picture}
}
=  \sum_{k}T[^k_j](u)T[^i_{k+1}](u)\,.
\end{eqnarray}
Also note the order in which operators $T$ stand. In general, $T$'s that are located to the left or down in the graphical notation correspond to operators that act first on the Hilbert space.

To define the transfer matrices in a representation $\lambda$, one uses fusion \cite{Zabrodin:1996vm}. Consider the following objects:
\begin{eqnarray}
\raisebox{-0.4\height}{
\begin{picture}(20,30)(0,-15)
\put(0,10){
\put(0,0){\line(1,0){3}}
\put(17,0){\line(1,0){3}}
}
\put(0,0){\line(1,0){3}}
\put(17,0){\line(1,0){3}}
\put(0,-10){
\put(0,0){\line(1,0){3}}
\put(17,0){\line(1,0){3}}
}
\polygon(3,14)(17,14)(17,-14)(3,-14)
\put(7,-2){$u$}
\end{picture}
}
\equiv
\raisebox{-0.4\height}{
\begin{picture}(44,30)(-3,-15)
\put(-3,0){\line(1,0){3}}
\put(-3,10){\line(1,0){3}}
\put(-3,-10){\line(1,0){3}}
\put(5,0){\oval(10,30)}
\put(2,-2){${\scriptstyle\wedge}$}
\put(10,0){
\put(0,10){
\put(0,0){\line(1,0){3}}
\put(31,0){\line(1,0){3}}
\polygon(3,4)(31,4)(31,-4)(3,-4)
\put(7,-2){${\scriptstyle u}$}
}
\put(0,0){
\put(0,0){\line(1,0){3}}
\put(31,0){\line(1,0){3}}
\polygon(3,4)(31,4)(31,-4)(3,-4)
\put(7,-2){${\scriptstyle u-\hbar}$}
}
\put(0,-10){
\put(0,0){\line(1,0){3}}
\put(31,0){\line(1,0){3}}
\polygon(3,4)(31,4)(31,-4)(3,-4)
\put(7,-2){${\scriptstyle\cdots}$}
}
}
\end{picture}
}
\,,\quad
\raisebox{-0.4\height}{
\begin{picture}(38,8)(-5,-4)
\put(-5,1){\line(1,0){8}}
\put(-5,-1){\line(1,0){8}}
\put(25,1){\line(1,0){8}}
\put(25,-1){\line(1,0){8}}
\polygon(3,4.5)(25,4.5)(25,-4.5)(3,-4.5)
\put(11,-2.5){$\scriptstyle \lambda$}
\end{picture}
}
\equiv
\raisebox{-0.4\height}{
\begin{picture}(44,30)(-3,-15)
\put(-3,0){\line(1,0){3}}
\put(-3,10){\line(1,0){3}}
\put(-3,-10){\line(1,0){3}}
\put(5,0){\oval(10,30)}
\put(2,-2){${\scriptstyle\Proj}$}
\put(10,0){
\put(0,10){
\put(0,0){\line(1,0){3}}
\put(31,0){\line(1,0){3}}
\polygon(3,4)(31,4)(31,-4)(3,-4)
\put(7,-2){${\scriptstyle v}$}
}
\put(0,0){
\put(0,0){\line(1,0){3}}
\put(31,0){\line(1,0){3}}
\polygon(3,4)(31,4)(31,-4)(3,-4)
\put(7,-2){${\scriptstyle v+\hbar}$}
}
\put(0,-10){
\put(0,0){\line(1,0){3}}
\put(31,0){\line(1,0){3}}
\polygon(3,4)(31,4)(31,-4)(3,-4)
\put(7,-2){${\scriptstyle \cdots}$}
}
}
\end{picture}
}\,.
\label{projdef}
\end{eqnarray}
The first one is the minor defined by \eqref{qminor} and here $\bigwedge$ stands for antisymmetrisation. The second object is the generalisation of the minor to the case of arbitrary representations. We will call it $\lambda$-minor or, equivalently, (a component of) the fused monodromy matrix. It is defined as follows. Consider the two Young tableaux of shape $\lambda$ populated by indices $\lA_{a,s}$ and $\lB_{a,s}$:
\begin{equation}
\lA=\raisebox{-0.4\height}{
\begin{picture}(90,70)(-10,0)
\put(-8,68){\vector(0,-1){30}}
\put(-8,68){\vector(1,0){40}}
\put(-16,42){$a$}
\put(25,72){$s$}
\drawline(0,0)(0,60)
\drawline(20,0)(20,60)
\drawline(40,20)(40,60)
\drawline(60,40)(60,60)
\drawline(80,40)(80,60)
\drawline(0,0)(20,0)
\drawline(0,20)(40,20)
\drawline(0,40)(80,40)
\drawline(0,60)(80,60)
\put(2,48){$\scriptstyle {\mathcal{A}}_{1,1}$}
\put(22,48){$\scriptstyle {\mathcal{A}}_{1,2}$}
\put(44,48){$\ldots$}
\put(60.5,48){$\scriptstyle {\mathcal{A}}_{1\!,\lambda_1}$}
\put(2,28){$\scriptstyle {\mathcal{A}}_{2,1}$}
\put(24,28){$\ldots$}
\put(4,8){$\ldots$}
\end{picture}
}\,,\quad
\lB=\raisebox{-0.4\height}{
\begin{picture}(90,70)(-10,0)
\drawline(0,0)(0,60)
\drawline(20,0)(20,60)
\drawline(40,20)(40,60)
\drawline(60,40)(60,60)
\drawline(80,40)(80,60)
\drawline(0,0)(20,0)
\drawline(0,20)(40,20)
\drawline(0,40)(80,40)
\drawline(0,60)(80,60)
\put(2,48){$\scriptstyle {\mathcal{B}}_{1,1}$}
\put(22,48){$\scriptstyle {\mathcal{B}}_{1,2}$}
\put(44,48){$\ldots$}
\put(60.5,48){$\scriptstyle {\mathcal{B}}_{1\!,\lambda_1}$}
\put(2,28){$\scriptstyle {\mathcal{B}}_{2,1}$}
\put(24,28){$\ldots$}
\put(4,8){$\ldots$}
\end{picture}
}\,.
\end{equation}
 Then explicitly
\begin{eqnarray}
\label{asminor}
T[^\lA_\lB][v]\equiv\sum_{\sigma,\tilde\sigma}\overrightarrow{\prod_{a=1}^{h_{\lambda}}\prod_{s=1}^{\lambda_a}}(-1)^{\tilde\sigma}T\left[^{\lA_{\tilde\sigma_s(a),\sigma_{\tilde\sigma_{s}(a)}(s)}}_{\lB_{a,s}}\right](v+\hbar(s-a))\,,
\end{eqnarray}
where $\sigma=\{\sigma_1\in\mathbb{S}_{\lambda_1},\sigma_2\in\mathbb{S}_{\lambda_2}\ldots\}$ and $\tilde\sigma=\{\tilde\sigma_1\in\mathbb{S}_{\lambda_1'},\tilde\sigma_2\in\mathbb{S}_{\lambda_2'},\ldots\}\,,$ $\lambda'$ denotes the transposed diagram and $h_\lambda=\lambda_1'$. The symmetrisation procedure \eqref{asminor} is the application of the Young symmetriser, explicitly written in indices, which projects $(\mathbb{C}^N)^{\otimes as}$ onto the irrep $\lambda$ of $\GL(N)$, and it is decoded as follows: first one symmetrises over the rows of the tableau $\mathcal{A}$, and then one anti-symmetrise over the columns. $\Proj$ in \eqref{projdef} stands for this symmetrisation procedure.

Although the symmetrisation is done explicitly only over the tableau $\lA$, the $\lambda$-minor has the same symmetry properties with respect to permutations of $\lB_{a,s}$. This is demonstrated by representing $\Proj$ as a properly regularised product of $R$-matrices with subsequent application of RTT relations, see {\it e.g.} \cite{Zabrodin:1996vm}. 

The transfer matrix $\wT_\lambda^G$ is defined by 
\begin{equation}
\wT_\lambda^G(u)= \sum_{\lA,\lB} \left(T[^\lA_\lB](u)\prod_{a,s}G_{{\lB_{a,s}},{\lA_{a,s}}}\right)\,,
\end{equation} where the sum is done over semi-standard Young tableaux. For the case of this subsection  $G$ is the null twist $\mathcal{N}$ and then $\wT_\lambda=\sum_{\lA}T[^\lA_{\lA+1}]\,.$

We will need scattering of the defining representation through the $\lambda$-minor:
\begin{eqnarray}
{
\raisebox{-0.4\height}{
\begin{picture}(50,30)(-3,-15)
\put(-3,0){\line(1,0){3}}
\put(-3,10){\line(1,0){3}}
\put(-3,-10){\line(1,0){3}}
\put(5,0){\oval(10,30)}
\put(2,-2){${\scriptstyle\Proj}$}
\put(10,0){
\put(0,10){
\put(0,0){\line(1,0){40}}
\put(4,2){${\scriptstyle v}$}
}
\put(0,0){
\put(0,0){\line(1,0){40}}
\put(4,2){${\scriptstyle v+\hbar}$}
}
\put(0,-10){
\put(0,0){\line(1,0){40}}
\put(4,2){${\scriptstyle \cdots}$}
}
}
\put(30,-20){\vector(1,3){14}}
\put(24,-18){$\scriptstyle u$}
\end{picture}
}
}
=\prod_{a,s}(v_{a,s}-u)\left(\Id+\frac{\hbar}{v-u}\Perm^{\lambda}\right)\,,
\label{scatter1}
\end{eqnarray}
where $\prod\limits_{a,s}f(v_{a,s})=\prod\limits_{a=1}^{h_\lambda}\prod\limits_{s=1}^{\lambda_a} f(v+\hbar(s-a))\,,$ and where $\Perm^{\lambda}$ is the generalised permutation \eqref{genPerm}.

Also we will need scattering in the opposite direction through an antisymmetric representation:
\begin{eqnarray}
{
\raisebox{-0.4\height}{
\begin{picture}(50,30)(-3,-15)
\put(-3,0){\line(1,0){3}}
\put(-3,10){\line(1,0){3}}
\put(-3,-10){\line(1,0){3}}
\put(5,0){\oval(10,30)}
\put(2,-2){${\scriptstyle\wedge}$}
\put(10,0){
\put(0,10){
\put(0,0){\line(1,0){40}}
\put(4,2){${\scriptstyle u}$}
}
\put(0,0){
\put(0,0){\line(1,0){40}}
\put(4,2){${\scriptstyle u-\hbar}$}
}
\put(0,-10){
\put(0,0){\line(1,0){40}}
\put(4,2){${\scriptstyle \cdots}$}
}
}
\put(30,20){\vector(1,-3){14}}
\put(32,16){$\scriptstyle v$}
\end{picture}
}
}
=\prod_{a=0}^{m-1}(v-u+\hbar a)\left(\Id+\frac{\hbar}{v-u+\hbar (m-1)\hbar}\Perm\right)\,.
\label{scatter2}
\end{eqnarray}

To introduce $\B$ in graphical notations, first recursively define $\B_k$, for $k=1,\ldots,N$, by
\begin{eqnarray}
\B_1\equiv 
\raisebox{-0.4\height}{
\begin{picture}(40,20)(0,-10)
\put(0,0){\line(1,0){13}}
\put(27,0){\line(1,0){13}}
\polygon(13,4)(27,4)(27,-4)(13,-4)
\put(17,-2){$u$}
\put(33,3){$\scriptstyle 1$}
\end{picture}
}\,,
\B_2\equiv \raisebox{-0.4\height}{
\begin{picture}(60,30)(0,-10)
\put(0,0){\line(1,0){3}}
\put(0,10){\line(1,0){3}}
\put(21,0){\line(1,0){10}}
\polygon(3,14)(21,14)(21,-4)(3,-4)
\put(4,4){$\scriptstyle u+\hbar$}
\put(18,0){
\put(27,0){\line(1,0){13}}
\polygon(13,4)(27,4)(27,-4)(13,-4)
\put(17,-2){$\scriptstyle u$}
\put(33,3){$\scriptstyle 1$}
}
\put(0,10){
\put(21,0){\line(1,0){13}}
\put(27,3){$\scriptstyle 1$}
}
\end{picture}
}\,,
\ldots\,,
\B_{k+1}\equiv \raisebox{-0.4\height}{
\begin{picture}(80,40)(0,-10)
\put(0,0){\line(1,0){3}}
\put(0,20){\line(1,0){3}}
\put(0,12){\line(1,0){3}}
\put(29,0){\line(1,0){8}}
\put(29,12){\line(1,0){8}}
\polygon(3,24)(29,24)(29,-4)(3,-4)
\put(6,8){$\scriptstyle u+k\hbar$}
\put(8,20){
\put(21,0){\line(1,0){13}}
\put(27,3){$\scriptstyle 1$}
}
\put(-4,0){
\polygon(41,-4)(41,16)(80,-4)
\put(48,0){$\scriptstyle B_{k}$}
}
\end{picture}
}\,.
\end{eqnarray}
Then $\B\propto\frac{1}{T[^{1\ldots N}_{1\ldots N}]}\B_{N}$, where we used that $T[^{1\ldots N}_{1\ldots N}]$ is the center of the Yangian.\\[1em]


\noindent Consider  the following chain of equalities
\begin{equation}
\begin{split}
\raisebox{-0.4\height}{
\begin{picture}(80,40)(0,-10)
\put(-10,0){\line(1,0){13}}
\put(-10,20){\line(1,0){13}}
\put(-10,12){\line(1,0){13}}
\put(29,0){\line(1,0){8}}
\put(29,12){\line(1,0){8}}
\polygon(3,24)(29,24)(29,-4)(3,-4)
\put(6,8){$\scriptstyle u+k\hbar$}
\put(8,20){
\put(21,0){\line(1,0){13}}
\put(27,3){$\scriptstyle 1$}
}
\put(-4,0){
\polygon(41,-4)(41,16)(80,-4)
\put(48,0){$\scriptstyle B_{k}$}
}
\put(0,-10){
\drawline(3,-1)(-5,-1)(-5,42)(85,42)(85,-1)(25,-1)
\drawline(3,1)(-3,1)(-3,40)(83,40)(83,1)(25,1)
\put(50,3){\line(5,-2){7.2}}
\put(50,-3){\line(5,2){7.2}}
\polygon(3,4.5)(25,4.5)(25,-4.5)(3,-4.5)
\put(11,-2.5){$\scriptstyle \lambda$}
}
\end{picture}
}
& =
\raisebox{-0.4\height}{
\begin{picture}(95,50)(-10,-10)
\put(-10,0){\line(1,0){13}}
\put(-10,20){\line(1,0){13}}
\put(-10,12){\line(1,0){13}}
\put(29,0){\line(1,0){8}}
\put(29,12){\line(1,0){8}}
\polygon(3,24)(29,24)(29,-4)(3,-4)
\put(6,8){$\scriptstyle u+k\hbar$}
\put(8,20){
\put(21,0){\line(1,0){13}}
\put(27,3){$\scriptstyle 1$}
}
\put(-4,0){
\polygon(41,-4)(41,16)(80,-4)
\put(48,0){$\scriptstyle B_{k}$}
}
\put(0,31){
\polygon(3,4.5)(25,4.5)(25,-4.5)(3,-4.5)
\put(11,-2.5){$\scriptstyle \lambda$}
}
\put(0,-10){
\drawline(3,40)(-5,40)(-5,49)(85,49)(85,-1)(32,-1)(32,40)(25,40)
\drawline(3,42)(-3,42)(-3,47)(83,47)(83,1)(34,1)(34,42)(25,42)
\put(50,3){\line(5,-2){7.2}}
\put(50,-3){\line(5,2){7.2}}
}
\end{picture}
}\\
& =
\prod_{a,s}(v_{a,s}-u-k\hbar)
\raisebox{-0.4\height}{
\begin{picture}(95,50)(-10,-10)
\put(-10,0){\line(1,0){13}}
\put(-10,26){\line(1,0){13}}
\put(-10,12){\line(1,0){13}}
\put(29,0){\line(1,0){8}}
\put(29,12){\line(1,0){8}}
\polygon(3,30)(29,30)(29,-4)(3,-4)
\put(6,8){$\scriptstyle u+k\hbar$}
\put(8,26){
\put(21,0){\line(1,0){13}}
\put(27,3){$\scriptstyle 1$}
}
\put(-4,0){
\polygon(41,-4)(41,16)(80,-4)
\put(48,0){$\scriptstyle B_{k}$}
}
\put(34,-10){
\polygon(3,4.5)(25,4.5)(25,-4.5)(3,-4.5)
\put(11,-2.5){$\scriptstyle \lambda$}
}
\put(0,-10){
\drawline(37,-1)(32,-1)(32,32)(82,32)(82,-1)(59,-1)
\drawline(37,1)(34,1)(34,30)(80,30)(80,1)(59,1)
\put(65,3){\line(5,-2){7.2}}
\put(65,-3){\line(5,2){7.2}}
}
\end{picture}
}
+\mathcal{R}_k(u,v)\,,\nonumber\\
\end{split}
\label{TBchain}
\end{equation}
where $\mathcal{R}_k=\sum_{j=1}^N \B_k(u)T_{j1}(v)\times\ldots\,,$ with dots standing for the expression whose explicit form is not relevant for further computations.

The first equality was obtained by applying the RTT relation between fused modromy matrices. It is easy to derive it directly from the definition \eqref{projdef} and repeated application of the elemental RTT relation \eqref{RTTgraph}. The important aspect is that the symmetry introduced by the Young symmetriser is present in the $\lB_{a,s}$ indices of $T[^\lA_\lB]$ and that it survives through scatterings.

The second equality was obtained from the following scattering, {\it cf.} \eqref{scatter1}, 
\begin{eqnarray}
\raisebox{-0.5\height}{
\begin{picture}(45,10)(-5,-8)
\put(-5,1){\line(1,0){8}}
\put(-5,-1){\line(1,0){8}}
\put(25,1){\line(1,0){18}}
\put(25,-1){\line(1,0){18}}
\polygon(3,4.5)(25,4.5)(25,-4.5)(3,-4.5)
\put(11,-2.5){$\scriptstyle \lambda$}
\put(30,-8){\vector(2,5){7}}
\put(37,9){$\scriptstyle 1$}
\end{picture}
}
=
\prod_{\alpha}(v_{\alpha}-u-k\hbar)
\raisebox{-0.5\height}{
\begin{picture}(45,10)(-5,-8)
\put(-5,1){\line(1,0){8}}
\put(-5,-1){\line(1,0){8}}
\put(25,1){\line(1,0){6}}
\put(25,-1){\line(1,0){6}}
\put(35,1){\line(1,0){8}}
\put(35,-1){\line(1,0){8}}
\polygon(3,4.5)(25,4.5)(25,-4.5)(3,-4.5)
\put(11,-2.5){$\scriptstyle \lambda$}
\put(30,-8){\vector(2,5){7}}
\put(37,9){$\scriptstyle 1$}
\end{picture}
}
+
\raisebox{-0.4\height}{
\begin{picture}(44,30)(-3,-15)
\put(-3,0){\line(1,0){3}}
\put(-3,10){\line(1,0){3}}
\put(-3,-10){\line(1,0){3}}
\put(5,0){\oval(10,30)}
\put(2,-2){${\scriptstyle\Proj}$}
\put(10,0){
\put(0,10){
\put(0,0){\line(1,0){3}}
\put(31,0){\line(1,0){5}}
\put(36,-2){$\scriptstyle 1$}
\polygon(3,4)(31,4)(31,-4)(3,-4)
\put(7,-2){${\scriptstyle v}$}
}
\put(0,0){
\put(0,0){\line(1,0){3}}
\put(31,0){\line(1,0){3}}
\polygon(3,4)(31,4)(31,-4)(3,-4)
\put(7,-2){${\scriptstyle v+\hbar}$}
}
\put(0,-10){
\put(0,0){\line(1,0){3}}
\put(31,0){\line(1,0){3}}
\polygon(3,4)(31,4)(31,-4)(3,-4)
\put(7,-2){${\scriptstyle \cdots}$}
}
}
\end{picture}
}
\times\ldots\,.
\label{eq332}
\end{eqnarray}
For the second term in the r.h.s., we need only that it is always of the form $\sum_{\lB}T[^\lA_{\lB}]\times\cdots$, where $1\in \lB$. By using the symmetry imposed by the Young symmetriser, one can prove that for any $\lB$ which contains $1$, one can represent the $\lambda$-minor as linear combination $T[^\lA_{\lB}]=\sum_{\lB'}\# T[^\lA_{\lB'}]$, where $\#$ stand for irrelevant for us numerical coefficients and all $\lB'$ are such that $\lB_{11}'=1$. To say it differently, the semi-standard Young tableaux label a basis of the irreducible representation $\lambda$, other Young tableaux are expressed through linear combinations of the semi-standard ones with the same content, and all semi-standard tableaux containing $1$ should have $\lB_{11}=1$. Then it follows that the second term in \eqref{eq332} is always of the form $\sum_j T[^j_1](v)\times\ldots$.

We use relatons \eqref{TBchain} to pull the trace over the $\lambda$-minors through the $\B$-operator. At the right-most step one gets $\wT_{\lambda}B_N$ plus $\mathcal{R}$-terms. At the left-most step, one uses the  scattering with the fully-antisymmetric representation
\begin{eqnarray}
{
\raisebox{-0.4\height}{
\begin{picture}(125,70)(0,-50)
\put(0,0){
\put(-3,0){\line(1,0){3}}
\put(-3,10){\line(1,0){3}}
\put(-3,-10){\line(1,0){3}}
\put(5,0){\oval(10,30)}
\put(2,-2){${\scriptstyle\Proj}$}
\put(10,0){
\put(0,10){
\put(0,0){\line(1,0){45}}
\put(4,2){${\scriptstyle v}$}
}
\put(0,0){
\put(0,0){\line(1,0){45}}
\put(4,2){${\scriptstyle v+\hbar}$}
}
\put(0,-10){
\put(0,0){\line(1,0){45}}
\put(4,2){${\scriptstyle \cdots}$}
}
}
}
\put(0,-35){
\put(-3,0){\line(1,0){3}}
\put(-3,10){\line(1,0){3}}
\put(-3,-10){\line(1,0){3}}
\put(5,0){\oval(10,30)}
\put(2,-2){${\scriptstyle\wedge}$}
\put(10,0){
\put(0,10){
\put(0,0){\line(1,0){45}}
\put(4,2){${\scriptstyle u+(N-1)\hbar}$}
}
\put(0,0){
\put(0,0){\line(1,0){45}}
\put(4,2){${\scriptstyle \cdots}$}
}
\put(0,-10){
\put(0,0){\line(1,0){45}}
\put(4,2){${\scriptstyle u}$}
}
}
}
\put(55,-10){\line(1,-1){35}}
\put(55,0){\line(1,-1){45}}
\put(55,10){\line(1,-1){55}}
\put(0,-35){
\put(55,10){\line(1,1){35}}
\put(55,0){\line(1,1){45}}
\put(55,-10){\line(1,1){55}}
}
\end{picture}
}
}
=\prod_{a,s}\frac{v_{a,s}-u+\hbar}{v_{a,s}-u}\prod_{k=0}^{N-1}(v_{a,s}-u-k\hbar)\times \Id\,
\end{eqnarray}
to take the trace cycle off the chain of $\B_k$'s hence producing an operator proportional to $\B\,\wT_{\lambda}$.

In summary, one gets the following relation
\begin{eqnarray}
\fbox{
$\displaystyle \wT_{\lambda}(v)\B(u)=\prod\limits_{a,s}\frac{u-v_{a,s}-\hbar}{u-v_{a,s}}\,\B(u)\wT_{\lambda}(v)+\mathcal{R}(u,v)\,,$
}
\label{TB}
\end{eqnarray}
where $\mathcal{R}(u,v)=\sum\limits_{k=0}^{N-1}\sum\limits_{j=1}^{N}\B_k(u)T_{j1}(v)\times\ldots\,,$ and the product over the Young tableau boxes reduces to the following explicit expression
\begin{eqnarray}
\prod\limits_{a,s}\frac{u-v_{a,s}-\hbar}{u-v_{a,s}}=\prod_{a=1}^{h_{\lambda}}\frac{u-v+\hbar\,(a-1-\lambda_a)}{u-v+\hbar\,(a-1)}\,.
\label{TB2}
\end{eqnarray}
To get use of \eqref{TB}, one needs to establish criteria when the remainder term $\mathcal{R}(u,v)$  vanishes. To this end we observe the following property. Suppose that some state $\bra{\Psi}$ satisfies $\bra{\Psi}T_{j1}(\theta)=0$ for all $j\in \{1,2,\ldots,N\}$ and some $\theta$. Then 
\begin{equation}\label{badvanish}
\bra{\Psi}\B_k(u)T_{j1}(\theta)=0\,
\end{equation}
for any $k,j$, whence $\bra{\Psi}\mathcal{R}(u,\theta)=0$. We will prove this in the next subsection. 

\subsection{Independence of twist eigenvalues}
We see that one needs to look for such states $\bra{\Psi}$ that $\bra{\Psi}T_{j1}(\theta)=0$ in order to apply (\ref{TB}) to the construction of $\B$-eigenstates. But these are also the states on which the action by transfer matrices evaluated at $\theta$ won't depend on twist! Indeed, from \eqref{twsitG} it follows that 
\begin{equation} 
\label{decomG}
T_{i,j}^G=T_{i,j+1}+T_{i,1}(-1)^{j-1}\chi_j\,,
\end{equation} and hence $\wT_{\lambda}^G=\wT_{\lambda}^{\lN}+\mathcal{R}$, where $\wT_{\lambda}^{\lN}$ is the null-twist transfer matrix and $\mathcal{R}=\sum_{\lB}T[^\lA_{\lB}]\times\cdots$ is the part depending on twist, with the property that all $\lB$ in the sum satisfy $1\in \lB$. By the symmetry argument as above one can write $\mathcal{R}(u)=\sum_j T_{j1}(u)\times\cdots$, and hence $\bra{\Psi}\wT_{\lambda}^G(\theta)=\bra{\Psi}\wT_{\lambda}^{\lN}(\theta)$ if $\bra{\Psi}T_{j1}(\theta)=0$.\\[0em]

Our main technical tool to work with the vectors of $\bra{\Psi}$-type and to generate new vectors of this type  will be the following commutation relation between a $\lambda$-minor $T\left[^\lA_\lB \right]$ and an element of the monodromy matrix
\begin{equation}\label{generalcomm}
\begin{split}
\frac{(v-u)}{\hbar}[T_{jk}(v),T\left[^\lA_\lB \right](u)] & =\sum_{i=1}^{|\lA|}T_{\lA_i k}(v) T\left[^{\lA[i;j]}_\lB \right](u) 
  -\sum_{i=1}^{|\lB|}T\left[^{\lA}_{\lB[i;k]} \right](u)T_{j \lB_i}(v)\,,
\end{split}
\end{equation}
where $\lA[i;j]$ denotes that the $i$-th entry of $\lA$ has been replaced by $j$, and the equivalent notation is used for $\lB$. The symmetry argument can be used to simplify $T\left[^{\lA}_{\lB[i;k]} \right](u)=\sum_{c\in\lA}T_{ck}(u)\times\dots\,$ so that \eqref{generalcomm} has the structure:
\begin{equation}
\label{simplification}
\begin{split}
(v-u)[T_{jk}(v),T\left[^\lA_\lB \right](u)] & =\sum_{c\in\lA}T_{ck}(v)\times\dots - \sum_{c\in\lA}T_{ck}(u)\times\dots\,.
\end{split}
\end{equation}

\noindent {\it Proof of \eqref{badvanish}.}  We demonstrate how to use \eqref{generalcomm}  by proving $(\ref{badvanish})$. In this particular case simplification \eqref{simplification} is used only for the $v$-term. 

Recall that each $\B_k$ is a sum of a product of (fully antisymmetric) quantum minors $T\left[^\lA_\lB \right]$, where $\lA,\lB \subset \{1,2,\dots,N\}$ and $1\in \lB$. We note that $\lB[i;1]=\delta_{i1}\lB$ due to antisymmetry, hence \eqref{generalcomm} is rewritten in this case as
\begin{equation}
\frac{(\theta-u)}{\hbar}T_{j1}(\theta)\,T\left[^\lA_\lB \right](u)-\frac{(\theta-u-\hbar)}{\hbar}T\left[^\lA_\lB \right](u)\,T_{j1}(\theta)= \sum_{c\in \lA}T_{c 1}(\theta)\times \dots \,.
\end{equation}
Then if $\bra{\Psi}T_{j1}(\theta)=0$ for all $j=1,\dots,N$, the above relation implies that $\bra{\Psi}T\left[^\lA_\lB \right](u)T_{j1}(\theta)=0$ thus proving \eqref{badvanish}. It seems that one needs to impose $u\neq\theta-\hbar$ to draw this conclusion, however all objects are polynomials in the rapidity variables; therefore if a quantity is zero on any dense set of $u$'s, it is identically zero, and the restriction $u\neq\theta-\hbar$ is unnecessary. \qed

\paragraph{Properties of the GT vacuum.} From now on we will restrict ourselves to the case of rectangular representations in the physical space that correspond to the rectangular Young diagram with $A\times S$ boxes. 

First we analyse the properties of the GT vacuum $\bra{0}$. This state has all $m_{kj}^{\alpha}$ introduced in \eqref{mintro} equal to zero, and also it is uniquely singled out by the fact that it is simultaneously an eigenvector of all the diagonal elements of the Yangian, with the following eigenvalues:
\begin{subequations}
\label{Yaneigenvalues}
\begin{align}
 \bra{0}T_{jj}(u) &= Q_\theta(u)\bra{0}, \quad j=1,\dots,N-A\,,
\\
 \bra{0}T_{jj}(u) &= Q^{[-2S]}_\theta(u)\bra{0}, \quad j=N-A+1,\dots,N\,.
\end{align}
\end{subequations}
Furthermore, since it is the dual of the lowest-weight state, we have
\begin{equation}
\bra{0}T_{kj}(u)=0,\quad 1\leq j< k\leq N\,.
\label{LWScondition}
\end{equation}
A slightly less obvious fact is that
\begin{equation}
\bra{0}T_{jk}(u)=0,\quad 1\leq j< k \leq N-A\,.
\label{shortening}
\end{equation}
This can be proved as follows:  any state different from $\bra{0}$, when expanded in the GT basis $\bra{\Lambda}^{\rm GT}$, would contain terms with excited (non-zero) values of $m_{N-A,1}^\alpha$ for at least some $\alpha=1,2,\ldots L$, as enforced by the branching rules \eqref{branchingrules}. On the other hand, the value of $m_{N-A,1}^\alpha$ can be read off from the eigenvalue of $T\left[^{12\dots N-A}_{12\dots N-A} \right](u)$ as was explained in Section 2. And then, since $T_{jk}$ for $1\leq j,k \leq N-A$ commutes with the latter minor, the action of $T_{jk}$ cannot affect the value of $m_{N-A,1}^{\alpha}$. Hence the action of $T_{jk}$ on $\bra{0}$ should be diagonal. Since $T_{jk}$, $j<k$ is nilpotent, all its eigenvalues are zero and thus \eqref{shortening} holds. We refer to \eqref{shortening} as the shortening condition as $\bra{0}T_{jk}(u)$ is not necessarily zero for more general classes of representations.

We combine some of the mentioned properties \eqref{Yaneigenvalues}-\eqref{shortening} into the following one
\begin{equation}
\label{eq:genshorten}
\bra{0}T_{jk}(u)=\delta_{jk}Q_\theta(u)\bra{0},\quad 1\leq k\leq N-A,\ j=1,\dots,N,
\end{equation}
which we will refer to as the generalised shortening condition. 

\paragraph{Recursive argument to prove twist independence.}
Now we shall act with transfer matrices $\T_\lambda(\theta_\alpha)$, $\alpha\in \{1,2,\dots,L\}$, on $\bra{0}$ and show that the result is indeed twist-independent provided that this action satisfies certain restrictions. To this end one proves the following 

\noindent {\it Proposition}: If $0\leq n< N-A-1$ and a vector $\bra{\Psi_{n+1}}$ satisfies the generalised shortening condition
\begin{equation}
\label{eq:genshortenn}
\bra{\Psi_{n+1}}T_{jk}(u)=\delta_{jk}Q_\theta(u)\bra{0},\quad 1\leq k\leq n+1,\ j=1,\dots,N,
\end{equation}
then  $\bra{\Psi_{n}}$ constructed as $\bra{\Psi_{n}}=\bra{\Psi_{n+1}}\prod\limits_{\alpha=1}^{L}\wT_{\lambda^{\alpha}}(\theta_{\alpha})$, for some collection of $L$ Young diagrams $\{\lambda^1,\lambda^2,\ldots,\lambda^L\}$, is also equal to $\bra{\Psi_{n}}=\bra{\Psi_{n+1}}\prod\limits_{\alpha=1}^{L}\wT_{\lambda^{\alpha}}^{\mathcal{N}}(\theta_{\alpha})$, and it  satisfies the generalised shortening \eqref{eq:genshortenn} with $n+1\to n$:
\begin{equation}
\label{eq:genshortenn2}
\bra{\Psi_{n}}T_{jk}(u)=\delta_{jk}Q_\theta(u)\bra{0},\quad 1\leq k\leq n,\ j=1,\dots,N\,.
\end{equation}
Moreover, for each such $\alpha$ that $\lambda^{\alpha}=\emptyset$ one has
\begin{equation}
\label{psin}
\bra{\Psi_{n}}T_{jk}(\theta_\alpha)=0,\quad 1\leq k\leq n+1,\ j=1,\dots,N\,.
\end{equation}
\noindent {\it Proof}: Consider
\begin{eqnarray}
\label{eq:some123}
\bra{\Psi_{n+1}}\T_\lambda(\theta_\alpha)&=&\bra{\Psi_{n+1}}\left(\sum_{\lA} T\left[^\lA_{\lA+1} \right](\theta_\alpha) + \sum_{j=1}^{N} \chi_j T_{j1}(\theta_\alpha)\times \dots \right)
\nonumber\\
&=&
\sum_{\lA}\bra{\Psi_{n+1}}T\left[^\lA_{\lA+1} \right](\theta_\alpha)=\bra{\Psi_{n+1}}\T_\lambda^{\mathcal{N}}(\theta_\alpha)\,,
\end{eqnarray}
as the $\chi_j$ terms vanish by the generalised shortening condition.
We now note  that the indices in the set $\lA$ can only take the values $n+1,n+2,\ldots,N$. Indeed, for any   $k\in \lA+1$ that satisfies $2\leq k\leq n+1$, one can write  $\bra{\Psi_{n+1}}T\left[^\lA_{\lA+1} \right](\theta_\alpha)=\sum_{j\in \lA}\bra{\Psi_{n+1}}T_{jk}(\theta_\alpha)\times \dots $
which  vanishes by the generalised shortening. Using (\ref{generalcomm}) it then follows that
\begin{equation}
\label{eq:some124}
\bra{\Psi_{n+1}}\T_\lambda(\theta_\alpha)T_{jk}(u)=\delta_{jk}Q_\theta(u)\bra{\Psi
_{n+1}}\T_\lambda(\theta_\alpha),\quad 1\leq k \leq n\,,
\end{equation}
where it was used $\bra{\Psi_{n+1}}T_{ck}(u)=0$ as $c\geq n+1$ (for the reason explained above) and $k\leq n$ (it was our choice used in \eqref{eq:some124}).

The commutation \eqref{generalcomm} can also be fruitfully used for $k=n+1$ case if $u\neq \theta_{\alpha}$, allowing us to demonstrate
\begin{equation}
\label{eq:some125}
\bra{\Psi_{n+1}}\T_\lambda(\theta_\alpha)T_{jk}(\theta_\beta)=0\quad 1\leq k \leq n+1\,,\quad\beta\neq\alpha\,.
\end{equation}
One now considers $\left[\bra{\Psi_{n+1}}\T_\lambda(\theta_\alpha)\right]\T_{\lambda'}(\theta_\beta)$ and the logic is repeated: $\left[\bra{\Psi_{n+1}}\T_\lambda(\theta_\alpha)\right]\T_{\lambda'}(\theta_\beta)=\left[\bra{\Psi_{n+1}}\T_\lambda(\theta_\alpha)\right]\sum_{\lA}T\left[^\lA_{\lA+1}\right](\theta_\beta)$ is concluded from \eqref{eq:some125} specialised to $k=1$. The fact that the set $\lA$ is restricted to take the values from $n+1,\ldots,N$ also follows from \eqref{eq:some125}, and then generalised shortening \eqref{eq:some124} for the state $\bra{\Psi_{n+1}}\T_\lambda(\theta_\alpha)\T_{\lambda'}(\theta_\beta)$ is obtained. A version of \eqref{eq:some125} for this state is derived in the same manner from \eqref{generalcomm}. Now it reads $\bra{0}\T_\lambda(\theta_\alpha)\T_\lambda'(\theta_\beta)T_{jk}(\theta_\gamma)=0$ for $1\leq k \leq n+1\,,\gamma\neq\alpha,\beta\,.$

We recursively run through all possible $\alpha$, at each loop getting $\bra{\Psi_{n+1}}\prod_{\alpha\in\mathcal{I}}\T_{\lambda^{\alpha}}(\theta_{\alpha})=\bra{\Psi_{n+1}}\prod_{\alpha\in\mathcal{I}}\T_{\lambda^{\alpha}}^{\mathcal{N}}(\theta_{\alpha})$, and updating \eqref{eq:some124} to
\begin{equation}
\label{eq:some124a}
\bra{\Psi_{n+1}}\prod_{\alpha\in\mathcal{I}}\T_{\lambda^{\alpha}}(\theta_{\alpha})T_{jk}(u)=\delta_{jk}Q_\theta(u)\bra{\Psi_{n+1}}\prod_{\alpha\in\mathcal{I}}\T_{\lambda^{\alpha}}(\theta_{\alpha}),\quad 1\leq k \leq n\,,
\end{equation}
 and \eqref{eq:some125} to
\begin{equation} 
\label{eq:some126}
\bra{\Psi_{n+1}}\prod_{\alpha\in\mathcal{I}}\T_{\lambda^{\alpha}}(\theta_\alpha)T_{jk}(\theta_\beta)=0\quad 1\leq k \leq n+1\,,\beta\notin \mathcal{I}\,,
\end{equation}
where $\mathcal{I}\subset\{1,2,\ldots,L\}$ is the set of $\alpha$'s that were used. Note that $n\geq 0$ and hence equation \eqref{eq:some126} is always valid for $k=1$ which is enough to show that the twist dependence drops from $\left[\bra{\Psi_{n+1}}\prod_{\alpha\in\mathcal{I}}\T_{\lambda^{\alpha}}(\theta_\alpha)\right]\T_{\lambda'}(\theta_{\beta})$.

When all possible values of $\alpha$ are exausted, relation \eqref{eq:some124a} becomes \eqref{eq:genshortenn2} and relation \eqref{eq:some126} becomes \eqref{psin}.
\qed

We now select $\bra{\Psi_{N-A}}\equiv\bra{0}$ as the starting point for the recursive application of the just proven theorem. One gradually decreases $n$ until we reach $\bra{\Psi_0}$, for which the generalised shortening \eqref{eq:genshortenn2} is an empty statement. However, at this stage we have already generated a multitude of different states sufficient for our goals.

The final expression for $\bra{\Psi_0}$ can be written as the following product
\begin{equation}
\bra{0}\prod_{n=1}^{N-A}\prod_{\alpha=1}^L \T_{\lambda^\alpha_n}(\theta_\alpha)\,,
\label{fresult}
\end{equation}
where $\lambda_n^{\alpha}$ are Young diagrams. The states of type $\bra{\Psi_{n}}$ can be represented by this product as well, one just needs to put some  $\lambda_n^{\alpha}=\emptyset$.

We have proven that the  product \eqref{fresult} is independent of the twist eigenvalues and hence is identical to the state created using null-twist transfer matrices. 

We have also proven that if one applies $\B$ to this state the remainder term $\mathcal{R}(u,\theta)$ at each step of commutation \eqref{TB} will vanish. It is not difficult to check that $\bra{0}{\rm Nil}_C=0$ and hence $\bra{0}\B=\bra{0}\B^{\rm GT}$. Therefore $\bra{0}$ is an eigenvector of $\B$ as it is an eigenvector of $\B^{\rm GT}$. Hence \eqref{fresult} is an eigenvector of $\B$, with known eigenvalues, thanks to \eqref{TB}, that will be explicitly written down below.

\subsection{The $\bra{\Lambda}$-basis}
By now we have established that \eqref{fresult} are eigenvectors of $\B$. Prior to interpreting them as basis vectors $\bra{\Lambda}$ \eqref{Xegvalues}, we should make several comments about their linear independence.

Transfer matrices can be represented as Wronskian-type determinants of Q-operators \cite{Krichever:1996qd,Tsuboi:2009ud,Bazhanov:2010jq,Kazakov:2010iu}:
\begin{eqnarray}
\label{Wronsk}
\wT_\lambda(u)=\frac{1}{Q_{\fullset}(u)}\det_{1\leq i,j\leq N}Q_i(u+\hbar\,\hat\lambda_j)\,,\quad\quad\hat\lambda_j=\lambda_j-j+1\,.
\end{eqnarray}
Then, by using Plucker identities, one can establish relations between products of $\wT$'s thus  not all \eqref{fresult} are linearly independent. The linearly independent subset can be constructed in multitude of ways, a possible choice is to restrict \eqref{fresult} to the products of transfer matrices for which the order $\lambda_n^\alpha\geq \lambda_{n+1}^\alpha$ can be established, where $\lambda\geq \lambda'$ means $\lambda_i\geq \lambda_i$ for all $i$.

There are no other possible relations between $\wT_{\lambda}$'s solely based on \eqref{Wronsk}. However evaluating $\wT_{\lambda}(u)$ at inhomogeneities of the spin chain poses further constraints. As it will follow from the computations in the next subsection,  $\wT_{\lambda}(\theta)$ is  non-zero, for generic values of $\theta_{\alpha}$ and $z_i$, as long as the Young diagram $\lambda$ is of height at most $A$ and of width at most $S$. Otherwise, $\wT_{\lambda}(\theta)=0$.

After the mentioned restrictions, the remaining vectors \eqref{fresult} become
\begin{eqnarray}
\label{eq:Lambdadef}
\fbox{
$\displaystyle
\bra{\Lambda}=\bra{0}\prod_{k=1}^{N-A}\prod_{\alpha=1}^L \T_{\mu^\alpha_k}(\theta_\alpha)\,,
$
}
\end{eqnarray}
where $\mu^{\alpha}_k=[m_{k1}^{\alpha}\geq \ldots\geq m_{kA}^{\alpha}]$ with $m_{kj}^{\alpha}$ being integers in the range $0,1,2,\ldots ,S$. Furthermore, the collection $\mu^{\alpha}_1,\ldots,\mu^{\alpha}_{N-A}$ should form an ordered set of Young diagrams, where for a pair of diagrams the bigger diagram is defined as the one that contains the other.

We now assume that the proposed set of $\bra{\Lambda}$'s is linearly independent and then, by counting, we get that it forms a basis in the Hilbert space. We will discuss the grounds for this assumption after we construct the wave functions in the next subsection. 

Since for $(S^A )$ representations a large section of the GT patterns is fixed, see Figure~\ref{fig:diamond2}, this results in a number of the operatorial roots of $\B$ simply being scalar multiples of the identity. This scalar part of $\B$ will be denoted by $\beta$, while the dynamical part will be denoted by $\mathbbm{b}$, so that $\B(u)=\beta(u)\mathbbm{b}(u)$ and
\begin{equation}
\beta(u)=\prod_{\alpha=1}^L\prod_{i=1}^{A-1}\prod_{j=1}^i(u-\theta_\alpha-\hbar\ (S-j+1))\prod_{i=1}^{N-A-1}\prod_{j=1}^i(u-\theta_\alpha+\hbar\ (A+j-1))\,,\end{equation}
\begin{equation}
\mathbbm{b}(u)=\prod_{\alpha=1}^L\prod_{k=1}^{N-A}\prod_{j=1}^A(u-\bx_{kj}^\alpha)\,.
\label{bsmall}
\end{equation}
Now it is an elementary exercise to apply the commutation relations \eqref{TB} for concluding that $\bra{\Lambda}$ are eigenstates of $\mathbbm{b}$ with eigenvalues
\begin{equation}
\bra{\Lambda}\bx_{kj}^{\alpha}=x_{kj}^{\alpha}\bra{\Lambda}\,,\quad x_{kj}^{\alpha}=\theta_\alpha+\hbar\,\hat m_{kj}^{\alpha}\,,\quad\hat m_{kj}^{\alpha}=m_{kj}^{\alpha}+1-j\,,
\end{equation}
so $m_{kj}^{\alpha}$ are nothing else but the elements of the GT pattern \eqref{mintro}. In the following we will use, depending on the context,  either $x$, $\hat m$, or $m$ to appropriately refer to the eigenvalues of $\bx$.

The construction of the $\bra{\Lambda}$-basis was done in a way that did not rely on the operator $\B$, and in particular $\bra{\Lambda}$'s do not depend on $u$. Hence the issues discussed at the end of section~\ref{sec:Spectrum} are resolved.

\subsection{Wave functions}
Suppose there is a function $\Psi(\bx)$ of operators $\bx_{kj}^{\alpha}$ such that $\braket{\Lambda|\tau}=\bra{\Lambda}\Psi(\bx)\ket{\Omega}$ for some eigenstate $\ket{\tau}$ of the Bethe algebra and for some reference state $\ket{\Omega}$ which is also an eigenstate. Let us compute this pairing in two different ways:
\begin{equation}
\bra{\Lambda}\tau\rangle=\bra{\Lambda}\Psi(\bx)\ket\Omega=\Psi(\theta+\hbar\ \hat{m})\bra{\Lambda}\Omega\rangle=
\Psi(\theta+\hbar\ \hat{m})\prod_{\alpha=1}^L\prod_{k=1}^{N-A} \T_{\mu^\alpha_k}^{\Omega}(\theta_\alpha)\bra{0}\Omega\rangle\,,
\end{equation}
\begin{equation}
\bra{\Lambda}\tau\rangle=\prod_{\alpha=1}^L\prod_{k=1}^{N-A} \T_{\mu^\alpha_k}^{\tau}(\theta_\alpha)\bra{0}\tau\rangle=
\Psi(\theta)\prod_{\alpha=1}^L\prod_{k=1}^{N-A} \T_{\mu^\alpha_k}^{\tau}(\theta_\alpha)\bra{0}\Omega\rangle\,,
\end{equation}
where superscripts $^\tau,^\Omega$ denote that one takes eigenvalues of the transfer matrices on corresponding Bethe states.

Assuming that the overlap $\braket{0|\Omega}\neq 0$, $\Psi$ should satisfy the following equation
\begin{equation}
\Psi(\theta+\hbar\ \hat{m})\prod_{k=1}^{N-A}\prod_{\alpha=1}^L \T_{\mu^\alpha_k}^{\Omega}(\theta_\alpha)=\Psi(\theta)\prod_{k=1}^{N-A}\prod_{\alpha=1}^L \T_{\mu^\alpha_k}^{\tau}(\theta_\alpha)\,
\label{WaveBig}
\end{equation}
that naturally suggests separation of variables. Indeed, for $\psi(x_1,\ldots,x_A)$ satisfying
\begin{equation}
\psi(\theta+m_j-j+1)\,\wT_{\mu}^{\Omega}(\theta)=\psi(\theta-j+1)\,\wT_{\mu}^{\tau}(\theta)
\label{waveeqn}
\end{equation}
for an arbitrary partition $\mu=[m_1,m_2,\ldots ,m_A]$  and inhomogeneity $\theta=\theta_{\alpha}\,, \alpha=1,\ldots,L$, one finds that
\begin{equation}
\Psi=\prod_{\alpha=1}^L\prod_{k=1}^{N-A}\psi(x_{k1}^{\alpha},\ldots,x_{kA}^{\alpha})\,
\end{equation}
solves \eqref{WaveBig}.

Note that while the variables $\bx_{kj}^\alpha$ with different $\alpha$ do factorise, the variables  with the same $\alpha$ do not, including the product $\prod\limits_{k=1}^{N-A}$ which is a spurious factorisation due to the constraint $m_{kj}\geq m_{k-1,j}$ on the admissible eigenvalues. However, this should be expected as eigenvalues of $X_{kj}^{\alpha}$ within the same GT pattern are constrained by one another. 

To solve \eqref{waveeqn}, we will use the Wronskian formula \eqref{Wronsk} supplemented with the important fact:
\begin{equation}
\label{onshellcondition}
Q_{\fullset}\equiv \det_{1\leq i,j\leq N}Q_i(u+\hbar\,(1-j))
\end{equation}
is an element of the center of the Yangian. Indeed, from \eqref{Wronsk} it follows that
\begin{equation}
\label{Qfulsetdefinition}
\frac{Q_{\fullset}^{[2]}}{Q_{\fullset}}=\wT_{N,1}=\det G\times T\left[^{12\ldots N}_{12\ldots N}\right]=\prod_{i=1}^L (z_iQ_{\theta}^{[-2(\hat\lambda_i+N-1)]})\times \Id\,.
\end{equation}

Then 
\begin{equation}
{
\psi(x)=\frac{\det\limits_{1\leq i,j\leq N}Q_i^{\tau}(x_j)}{\det\limits_{1\leq i,j\leq N}Q_i^{\Omega}(x_j)}\,,
}
\label{psi1}
\end{equation}
where one sets $x_j \equiv \theta+\hbar(1-j)$  for $j>A$.

In the following the abridged notation shall be used: $x\equiv (x_1,\ldots,x_A)$ stands for  $x^{\alpha}_k\equiv (x^{\alpha}_{k1},\ldots, x^{\alpha}_{kA})$, for some $\alpha,k$. \\[0em]

Let us discuss now the structure of the eigenvalues of $Q$-operators in more detail.

There is a family of $2^N$ Q-operators labelled by multi-indices from $\{1,2,\ldots,N\}$ that were explicitly constructed for rational $\gl(N)$ spin chains in a wide class of representations \cite{Bazhanov:2010jq,Kazakov:2010iu,Frassek:2011aa}, in particular for compact ones - see for example \cite{Bazhanov:1996dr,Derkachov:2003qb,Tsuboi:2009ud,Niccoli:2010sh} for the construction of Q-operators in various other contexts. Hence the requested structure of  the Q-eigenvalues is explicitly known. With adjustments to our conventions the Q-eigenvalues should be of the form
\begin{equation}
Q_{I}\propto \hat q_I\, \times\,\Gamma\left[\prod_{j=1}^{|I|}Q_{\theta}^{[-2(\hat\lambda_j+
|I|-1)]}\, \right]\,,\quad \hat q_I\equiv q_I\times \prod_{i\in I}z_{i}^{\frac{u}{\hbar}}\,,
\label{Q1}
\end{equation}
where $q_I$ is a monic polynomial and $\hat q_I$ is dubbed twisted polynomial. This result can be obtained either by direct construction or by analysing the structure of solutions to Hirota equations \cite{Kazakov:2007fy,Kazakov:2015efa}. The function ${\Gamma}[F]$ is defined by its property ${\Gamma}[F^{[2]}]=F\,{\Gamma}[F]$. Although ${\Gamma}[F]$  is only fixed up to multiplication with an $\hbar$-periodic function, such a function cancels out in any expressions of direct interest for us. In the following $Q$, $q$ will refer to the eigenvalues of Q-operators dubbed Q-functions, unless stated otherwise. Superscripts $^{\tau,\Omega}$ will be dropped unless this leads to ambiguous expressions.

The Q-functions should satisfy the QQ-relations
\begin{equation}
\label{QQ}
Q_{Iij}Q_{I}^{[-2]}\propto Q_{Ii}Q_{Ij}^{[-2]}-Q_{Ij}Q_{Ii}^{[-2]}\,,
\end{equation}
see {\it e.g.} \cite{Kazakov:2007fy,Tsuboi:2009ud,Frassek:2011aa} and references therein. These relations, together with the requirement $q_{\fullset}=1$, {\it cf.} \eqref{Qfulsetdefinition}, restricts the possible values of $q_I$. For the case of the defining representation of the spin chain sites, this set is discrete and was proven \cite{2013arXiv1303.1578M} to be in one-to-one correspondence with the eigenstates of the Bethe algebra.  It is very likely that the same statement holds for any $(S^A)$ with $S=1$, while for $S>1$ one should also demand that transfer matrices evaluated by \eqref{Wronsk} do not have poles at $u=\theta_{\alpha}+\hbar\,\mathbb{Z}$. This extra requirement is needed to distinguish $(S^A)$ length $L$ spin chains from $(A\times 1)$ length $L\times S$ spin chains with special arrangement of inhomogeneities.


Although one can recast \eqref{QQ} into conventional Bethe equations, see {\it e.g.} \cite{Kazakov:2007fy,Tsuboi:2009ud,Frassek:2011aa},  \eqref{QQ} proves to be more efficient tool for some applications \cite{Marboe:2016yyn},  and probably it is the most reasonable one in more complicated systems where polynomiality is absent, like in the case of AdS/CFT integrability \cite{Gromov:2014caa}.\\[0em]

We will now use \eqref{Q1} and \eqref{QQ} to further simplify our expression for the wave function \eqref{psi1}. Start by noting that
\begin{equation}
\det_{1\leq i,j\leq N} Q_i(x_j)=\det_{1\leq i,j\leq N} \hat q_i(x_j)\times \Gamma\left[\prod_{j=1}^{N}Q_{\theta}(x_j-\hbar\,S)\, \right]\,.
\end{equation}
Denote by $\hat q_{(k)}$ the rank-$k$ skew-symmetric tensors whose components are $\hat q_{I}$ with $|I|=k$. Then, first (since a rank-$N$ tensor has only one component we identify it with scalar function)
\begin{equation}
\det_{1\leq i,j\leq N} \hat q_i(x_j)=\bigwedge_{j=1}^A\hat q_{(1)}(x_j)\wedge\bigwedge_{j=A+1}^{N} \hat q_{(1)}(\theta+\hbar\,(1-j))\,;
\end{equation}
and, second, one of the consequences of $\eqref{QQ}$ and $\eqref{Q1}$ is
\begin{equation}
\hat q_{(1)}\wedge \hat q_{(1)}^{[-2]}\wedge\ldots \hat q_{(1)}^{[-2A]}=\hat q_{(A+1)}\times Q_{\theta}^{[-2]}\ldots Q_{\theta}^{[-2S]}\,.
\end{equation}
The r.h.s. of the last equation obviously vanishes for $u=\theta+\hbar,\theta+2\hbar,\ldots,\theta+S\,\hbar$ which implies that any set of $A+1$ distinct vectors $\hat q_{(1)}(\theta+r\,\hbar)$ with $-A+1\leq r\leq S$ is linearly dependent. Given that the eigenvalues $x_j$ fall into the mentioned range of $\theta+r\,\hbar$, we conclude that $\hat q_{(1)}(x_j)$  can be always expressed as a linear combination of $\hat q_{(1)}(\theta+r'\,\hbar)$ for $r'=0,1,\ldots,A-1$, and therefore
\begin{equation}
\label{tensorequality}
\bigwedge\limits_{j=1}^A\hat q_{(1)}(x_j)=C(x)\,\bigwedge_{j=1}^{A} \hat q_{(1)}(\theta+\hbar\,(1-j))\,.
\end{equation}
Note also that $\bigwedge\limits_{j=1}^{A+1}\hat q_{(1)}(x_j)=0$ if $x_{A+1}=\theta+\hbar(m-A)$, for some positive integer $m$. This implies that $T_{\mu}(\theta)$ identically vanishes if the height of the Young diagram $\mu$ exceeds $A$, which is consistent with the fact that we restrict ourselves to only $A$ variables $\bx$.

The coefficient of proportionality $C(x)$ can be read off from some component of the tensor equality \eqref{tensorequality}:
\begin{equation}
C(x)=\frac{\det\limits_{i\in I,1\leq j\leq A} \hat q_i(x_j)}{\det\limits_{i\in I,1\leq j\leq A} \hat q_i(\theta-\hbar(j-1))}\,.
\end{equation}
We emphasise that $I$ can be any length-$A$ subset of $\{1,2,\ldots,N\}$.  Until the end of the section, for simplicity of notation, we choose $I=\{1,2,\ldots,A\}$.

We use the obtained results to conclude, after some reorganisation, that 
\begin{equation}
\det\limits_{1\leq i,j\leq N} Q_i(x_j)=\frac{\det\limits_{1\leq i,j\leq A}\hat q_i(x_j)}{\hat q_{12\ldots A}(\theta)}\,Q_{\fullset}(\theta)\,\Phi(x)\,,\quad \Phi(x)=\prod_{j=1}^A\prod_{r=1}^{m_j}Q_\theta(\theta-\hbar(S+j-r))\,.
\label{detQ3}
\end{equation}
The factors $Q_{\fullset}(\theta)$ and $\Phi$ are universal,  they do not depend on the choice of the Bethe algebra eigenvector. We remind the reader that the spectrum of the operator $X_j$ is $x_j=\theta+m_j-j+1$, $0\leq m_j\leq S$, and the upper bound on $m_j$ is consistent with \eqref{detQ3}.  Indeed,  $\Phi(x)$ vanishes if some $m_j$ exceeds $S$ which implies that $T_{\mu}(\theta)$ identically vanishes if the width of the Young diagram $\mu$ exceeds $S$. \\[0em]

Next, one turns to introducing  a convenient reference state $\ket{\Omega}$. In the frame where the twist is diagonal, the highest-weight vector of the Yangian representation, an analog of a ferromagnetic vacuum, is an eigenstate of the Bethe algebra. By applying Weyl symmetries of $\GL(N)$, which are permutations $\sigma\in\mathbb{S}_N$, we can obtain in total $\frac{N!}{A!(N-A)!}$ different states, all with particularly simple properties. By $\ket{\Omega_\sigma}$ we denote such states rotated to the companion twist basis and normalised to $\bra{0}\Omega_{\sigma}\rangle=1$. The fact that $\bra{0}\Omega_{\sigma}\rangle\neq 0$ and hence such a normalisation is possible will be proven shortly below.

 The reference states $\ket{\Omega_\sigma}$ satisfy the following property
\begin{equation}
T_{jj}^G \ket{\Omega_{\sigma}} = z_{j}Q_{\theta}^{[-2\lambda_{\sigma(j)}]}\ket{\Omega_{\sigma}}\,,
\end{equation}
and hence $\wT_{1,1}^{\Omega_{\sigma}}(\theta)\ket{\Omega_{\sigma}} =\left(\sum\limits_{i=1}^A z_{\sigma^{-1}(j)}\right)Q_{\theta}(\theta-\hbar\,S)\ket{\Omega_{\sigma}} $.  Likewise, eigenvalues of other transfer matrices evaluated at $\theta$ are also  expressible in terms of $Q_{\theta}$  and proportional to symmetric polynomials in $z_{\sigma^{-1}(1)},\ldots, z_{\sigma^{-1}(A)}$. The explicit expression for $\wT_{\mu}^{\Omega_\sigma}(\theta)$ follows from \eqref{detQ3} and \eqref{Wronsk} after one computes, either from the definition of Q-operators or by analysing QQ-relations,  $q_{\sigma(1)}^{\Omega_\sigma}=q_{\sigma(2)}^{\Omega_\sigma}=\ldots=q_{\sigma(A)}^{\Omega_\sigma}=1$ which implies
\begin{equation}
\frac{\wT_{\mu}^{\Omega_\sigma}(\theta)}{\Phi(x)}=\frac{\det\limits_{1\leq i,j\leq A}\hat q_i^{\Omega_\sigma}(x_j)}{\hat q^{\Omega_\sigma}_{12\ldots A}(\theta)}=\frac{\det\limits_{1\leq i,j\leq A} z_{\sigma^{-1}(i)}^{m_j-j+1}}{\det\limits_{1\leq i,j\leq A} z_{\sigma^{-1}(i)}^{-j+1}}=\chi_{\mu}(z^{\sigma})\,,
\label{detQ4}
\end{equation}
where $(z^{\sigma})_i\equiv z_{\sigma^{-1}(i)}$.

Hence it would be natural to introduce the normalised SV basis
\begin{equation}
\label{xnormalised}
\bra{{\bf x}}\equiv \bra{\Lambda}\prod_{\alpha,k}\frac 1{\Phi(X_k^{\alpha})}\,,\quad \text{so that}\quad \bra{{\bf x}}\Omega_{\sigma}\rangle=\prod_{\alpha,k}\chi_{\mu_{k}^{\alpha}}(z^{\sigma})\,.
\end{equation}

From \eqref{detQ3} and \eqref{detQ4} we conclude that \eqref{psi1} simplifies to  
\begin{equation}\psi(x)=\frac 1{\chi_{\mu}(z^{\sigma})} \frac{\det\limits_{1\leq i,j\leq A}\hat q_i^{\tau}(x_j)}{\hat q_{12\ldots A}^{\tau}(\theta)}\,
\end{equation}
and hence the eigenvectors $\ket{\tau}$ of the Bethe algebra are constructed as
\begin{equation}
\label{generationofstates}
\fbox{
$\displaystyle
\ket{\tau}=\prod_{\alpha=1}^{L}\prod_{k=1}^{N-A}\frac{\det\limits_{i\in I,1\leq j\leq A}\hat q_i(X_{kj}^{\alpha})}{\chi_{\mu_{k}^{\alpha}}(z^{\sigma})\hat q_{I}(\theta)}\ket{\Omega_{\sigma}}\,,
$
}
\end{equation}
where $\sigma$ can be arbitrary and where we also restored arbitrariness in the choice of $I$.

The  $\hat q_i$ are "on-shell"  twisted polynomials solving the QQ-relations/Bethe equations. It would be natural to define "off-shell" Bethe vectors $\ket{\tau}$ by \eqref{generationofstates} with arbitrary polynomials $\hat q_i$, we however did not check how this compares to the off-shell Bethe vectors studied in the literature \cite{Hutsalyuk:2016srn,Liashyk:2018egk}.

Finally, we return to the question of linear independence of $\bra{\Lambda}$'s which is needed for the derivation of the obtained results. If the spectrum of $\B$ is non-degenerate, all  $\bra{\Lambda}$'s  have different $\B$-eigenvalues and the linear independence is clear provided $\bra{\Lambda}\neq 0$. To show non-vanishing of $\bra{\Lambda}$'s  we compute 

\begin{equation}
\bra{\Lambda}\Omega_{\sigma}\rangle=\prod_{\alpha,k} \T_{\mu^\alpha_k}^{\Omega_{\sigma}}(\theta_\alpha)\bra{0}\Omega_{\sigma}\rangle= \chi_{\mu}(z^{\sigma})\bra{0}\Omega_{\sigma}\rangle\,.
\end{equation}
Now recall that $\bra{0}=\bra{\rm LWS}^{\rm GT}$ and remark that, for $\sigma: i\mapsto N-i+1$, $\ket{\Omega_{\sigma}}=c\,\Pi(K)\ket{\rm LWS}^{\rm GT}$, where $c$ is some normalisation constant which is obviously non-zero and $K_{ij}=z_j^{N-i}$ performs the similarity transformation $G=KgK^{-1}$ between $g={\rm diag}(z_1,\ldots, z_N)$ and $G$ -- the companion matrix \eqref{eq:Gcompanion}.

 Explicitly, the lowest-weight state is the vector $\ket{\rm LWS}^{\rm GT}=\bigotimes\limits_{\alpha=1}^L(e_{N}^{(\alpha)}\wedge \ldots e_{N-A+1}^{(\alpha)})^{\otimes S}$, where $e_i^{(\alpha)}$ are the standard $\GL(N)$-covariant basis vectors of the $\alpha$-th site of the spin chain. Then we can explicitly compute
\begin{equation}
\bra{0}\Omega_{\sigma}\rangle=c\,\bra{\rm LWS}^{\rm GT}\Pi(K)\ket{\rm LWS}^{\rm GT}=c\,\left(\prod\limits_{N-A-1\leq  i<j\leq N}(z_{i}-z_{j})\right)^{L\times S}\neq 0\,,
\end{equation} 
and hence $\bra{\Lambda}\neq 0$. In this derivation we have chosen $z_i$ to be pair-wise distinct, but since $\bra{\Lambda}$ do not depend on twist eigenvalues the conclusion $\bra{\Lambda}\neq 0$ holds for any $z_i$.

If $\B$ is degenerate, we argue that the linear independence still holds. It is always true that
\begin{equation}
\label{eq:lasteq}
{\bra{\Lambda}\tau\rangle}=\prod_{\alpha,k} \T_{\mu^\alpha_k}^{\tau}(\theta_\alpha){\bra{0}\tau\rangle}\,.
\end{equation}
We can expect that  ${\bra{0}\tau\rangle}={\bra{\Omega^{\sigma'}}U|\tau\rangle}\neq 0$ for generic enough $z$'s, so $\bra{\Lambda}\neq 0$. Due to \eqref{eq:lasteq}, linear dependence between $\bra{\Lambda}$ translates into an equation between  $\T_{\mu^\alpha_k}^{\tau}(\theta_{\alpha})$. This equation should be polynomial in $\theta_\alpha,z_i$ and Bethe roots (zeros of the polynomials $q(u)$), and also $\theta_{\alpha},z_i$ and Bethe roots are connected by the QQ-relations which are polynomial. Hence if it holds generically, it should hold always, i.e. to be a relation between transfer matrices but not their particular eigenvalues. This cannot be a consequence of the Wronskian formula \eqref{Wronsk} and QQ-relations as those were already fully used. Therefore if one shows that the Bethe algebra is isomorphic to a ring generated by polynomials $q_i$ subject to TQ- and QQ-relations, new type of relations are impossible and hence $\bra{\Lambda}$ are linearly independent.

The question of isomorphism of the Bethe algebra to the mentioned ring is essentially the question of completeness. It is fully resolved for the defining representation \cite{2013arXiv1303.1578M} but remains an open question for spin chains in other representation, though the results of \cite{2013arXiv1303.1578M} are likely to be generalisable.

Using the fact that $\bra{\Lambda}$ do not depend on $z_i$ and by varying $z_i$ in \eqref{eq:lasteq} one can attempt to circumvent the question of completeness of the Bethe equations. For instance, the right equality in \eqref{xnormalised} allows one to distinguish many, but not all, $\bra{\Lambda}$. Similarly, taking the singular twist limit of the Bethe algebra allows one to distinguish many, but generically not all, $\bra{\Lambda}$. Using these observations combined we confirmed linear independence for $N\leq 5$ without appealing to completeness.

\section{Conclusions and final remarks}
In this work we accomplished several steps towards explicit realisation of the SoV program for rational $\gl(N)$ spin chains. The main result is the construction of such a separated variable basis $\bra{\bf x}$ independent of twist eigenvalues $z_i$  that the wave functions $\bra{\bf x}\tau\rangle $ of the Bethe vectors $\ket{\tau}$ in the companion twist frame factorise into product of Slater determinants:
\begin{equation}
\label{eq:finalresultforxt}
\fbox{
$\displaystyle
\langle {\bf x}\ket{\tau}=\left(\frac{1}{{\hat q_{I}(\theta)}}\right)^{L(N-A)}\prod_{\alpha=1}^{L}\prod_{k=1}^{N-A}{{\det\limits_{i\in I,1\leq j\leq A}\hat q_i(x_{kj}^{\alpha})}}\,,
$
}
\end{equation}
where $I$ is arbitrary length-$A$ subset of $\{1,\ldots,N\}$;  $x_{kj}^{\alpha}$ are the eigenvalues of the separated variables $X_{kj}^{\alpha}$ with  explicit spectrum given by
\begin{equation}
\bra{\bf x}\bx_{kj}^{\alpha}=x_{kj}^{\alpha}\bra{\bf x}\,,\quad x_{kj}^{\alpha}=\theta_\alpha+\hbar\,(m_{kj}^{\alpha}+1-j)\,,
\end{equation}
where integers $m_{kj}^{\alpha}$ should satisfy constraints $0\leq m_{k,j}^{\alpha}\leq m_{k,j-1}^{\alpha}\leq S$ and $m_{kj}^{\alpha}\leq m_{k+1,j}^{\alpha}$; and, finally, $\hat q_i=z_i^{\frac u{\hbar}}q_i$ with $q_i$ being the Baxter polynomials.

 There are  $N\choose A$ "ferromagnetic vacua" states $\ket{\Omega_{\sigma}}$, with
\begin{equation}
\label{eq:xO}
\langle {\bf x}\ket{\Omega_{\sigma}}=\prod_{\alpha=1}^L\prod_{k=1}^{N-A}\chi_{\mu_k^\alpha}(z_{\sigma^{-1}(1)},\ldots,z_{\sigma^{-1}(A)}),
\end{equation}
where $\sigma\in\mathbb{S}_N$ is an element of the $\gl(N)$ Weyl group. These vacua can be used as reference states to generate all the others using \eqref{generationofstates}. Note also that the SV basis possesses a special state $\bra{0}$ dubbed the GT vacuum which has the property $\langle{0}\ket{\tau}=1$.

The result is obtained for rational spin chains in arbitrary rectangular representations $(S^A)$ and for arbitrary rank $N$. We assumed that spin chain inhomogeneities are not equal and that they are in generic position to ensure that accidental degeneracies are not present. The twist eigenvalues are arbitrary and can in principle be equal to each other. However, if  $z_i=z_j$ for some $i,j$, the companion twist matrix \eqref{eq:Gcompanion} is no longer similar to a diagonal matrix and hence in such a case the obtained results cannot be rotated to frames where the twist is diagonal.

Most of the derivations are robust, save for the proposition that vectors $\bra{\Lambda}$ defined by \eqref{eq:Lambdadef} are linearly independent. This proposition is proven only for the case when $\B$ is non-degenerate, that is for the symmetric and anti-symmetric powers of the defining representation, and for their conjugates. When the spectrum of $\B$ is degenerate, the linear independence question is reduced, using general position arguments, to the completeness  of the Bethe equations statement which, although is likely to be true, is not a rigorously proven statement in the literature save for $A,S=1$. Hence this part of the exposition requires further study.

The $\hat q$-functions satisfy the Baxter equation which can be compactly written in the following form, {\it cf.} \eqref{YangCappeli} and see also \cite{Chervov:2007bb},
\begin{equation}
{\rm det}(1+T^G(u)Q_{\theta}^{[s]}e^{-\hbar \partial_u})\hat q_i^{[2]}(u)=0\,,\quad i=1,2,\ldots,N\,.
\end{equation}
Hence the $\hat q$-functions are indeed true wave functions in the sense that they satisfy a wave equation. 

The Baxter equation should be perceived as a quantisation of the classical spectral curve equation 
$\det(y-M(u))=0$ present in classical integrable systems with Lax connection. This point of view generalises the original ideas of Sklyanin \cite{10.1007/3-540-15213-X_80,Sklyanin:1995bm} to higher-rank cases. More accurately, the Baxter equation is not unique, but there are $L\times (N-A)\times A$ copies of it. At each copy, $u$ takes only a finite set of values -- the eigenvalues of $X_{kj}^\alpha$. On a classical level, these eigenvalues become positions $x_i$ of the dynamical divisor, and the equations $\det(e^{p_i}-M(x_i))=0$ allow one to find the canonical conjugate of $x$ in the sense of Poisson brackets $\{x_i,p_j\}=\delta_{ij}$.  We refer to \cite{babelon_bernard_talon_2003} for further details on the classical spectral curve.

The separated variables $X_{kj}^{\alpha}$ are those operatorial zeros of the operator $\B(u)$ that are non-constant for $(S^A)$ representations.  $\B(u)$ is equal to $\B^{K_1,K_2}=B_{\rm good}$ -- the operator proposed in the paper of Gromov, Levkovich-Maslyuk and Sizov \cite{Gromov:2016itr}  -- with  $K_1=\Id$ and $K_2=G$ being the companion twist. If one wants to use our results  for building eigenstates of the Bethe algebra $\mathcal{B}^g$ with a diagonal twist $g={\rm diag}(z_1,\ldots, z_N)$, one needs to rotate the physical space or, equivalently, to choose $K_1=K$, $K_2=K^{-1}G$, where $g=K^{-1}GK$, for instance $K_{ij}=z_j^{N-i}$ is a possible choice.

The operator $\B$ differs from $\B^{\rm GT}$ -- an element in the GT algebra -- only by a nilpotent operator which allowed us to explicitly find the eigenvalues of $\B$ that are labelled by GT patterns.

A special property of the companion twist frame is that $\B$ does not depend on twist eigenvalues. We were able to show that appropriate action of transfer matrices on the dual of the GT lowest-weight state $\bra{0}$ does not depend on them either and, by considering the null-twist to simplify computations, we showed that this action generates eigenstates of $\B$. Our attempt to find eigenstates through action by transfer matrices was inspired by the idea Maillet and Niccoli presented in \cite{Maillet:2018bim}. Note that eigenstates are typically generated not by systematically increasing the powers of transfer matrices, as was suggested in \cite{Maillet:2018bim}, but rather by increasing the representation in which these transfer matrices are computed, so taking powers is replaced by a more covariant operation -- fusion.

As eigenstates of separated variables are generated by transfer matrices, there is now a straight road to get to the Slater determinants by exploiting Wronskian expressions for the transfer matrices, which was the last step we performed to arrive to our final result \eqref{eq:finalresultforxt}. Using a Slater determinant instead of a direct product of Q-functions is essential as it allows one to exploit the on-shell determinant condition \eqref{onshellcondition} which is a consequence of Bethe equations/QQ-relations. \\[0em]

There are still certain questions of the SoV program that remain to be resolved.

First, we recall that $\bra{v}w\rangle$ was defined as natural pairing between a dual vector and a vector, but not as a scalar product. In fact, we never introduced any metric in this work. The computation of norms of Bethe vectors and {\it e.g.} scalar products between off-shell and on-shell Bethe vectors was not considered, but of course it is an important question that should be studied in the future. Such computation in the case of $\gla(2)$ spin chains in the SoV framework was previously considered in \citep{Kazama:2013rya,Kitanine:2015jna} and it would be very interesting to generalise these findings to higher rank.

Secondly, the construction of the ladder operators $P^{\pm}$ in an explicit way, probably similar to \eqref{ladderoperators},  is yet to be done. A related question is finding a framework where wave functions fully factorise, with no intertwining by the Slater determinant. These questions can be potentially resolved if we learn how to generalise our approach to more general classes of representations of the physical space. In particular it would be instructive to consider generic representations of $\GL(N)$ where no highest- or lowest-weight states are present, and with Cartan charges being non-integer. For these, the entries of GT patterns would read as $\lambda_{ij}=\theta_{ij}+\hbar\, n_{ij}$, where $\theta_{ij}$ are distinct complex numbers and $n_{ij}$ are integers with no upper or lower bounds on their values. Hence the eigenvalues of $X_{ij}$ will not be constrained by one another, therefore we expect wave functions to be directly given by products of Q-funcitons. Also the classical limit would be more accessible for these representations, as the representation space is infinite. Both the GT algebra of the Yangian \cite{Valinevich:2016cwq} and eigenvectors of $\B$ \cite{Derkachov:2018ewi} for $N=3$ were already explored in this case, so there is a good starting point.

        Finally, we remark that $\B(u)$ contains only symmetric polynomials in separated variables, while our wave functions have only part of this symmetry. Hence it seems to be not possible, in general, to generate eigenstates in the style $\prod_r \B(u_r)\ket{\Omega}$, where $u_r$ are some numbers, probably Bethe roots. One notable exception is spin chains in symmetric powers of the defining representation, i.e. with $A=1$. In this case only the variables $X_{k1}^{\alpha}$ are not constants, and \eqref{generationofstates} simplifies to
$\ket{\tau}=\prod\limits_{\alpha=1}^L \prod\limits_{k=1}^{N-1}\frac{\hat q_{i}(X_{k1}^{\alpha})}{z_{j}^{m_k}\hat q_i(\theta)}\ket{\Omega_j}$, for any $i,j=1,2,\ldots N$. Recall that $\hat q_i=z_i^{u/\hbar}q_i$. Hence further cancellations occur if one chooses $j=i$, and one gets
\begin{equation}
\ket{\tau}=\prod_{\alpha=1}^L \prod_{k=1}^{N-1}\frac{q_{i}(X_{k1}^{\alpha})}{q_i(\theta)}\ket{\Omega_i}\propto \prod_{r}\B(u_r^{(i)})\ket{\Omega_i}\,,
\end{equation}
where $u_r^{(i)}$ are zeros of $q_i(u)$. This is an explicit derivation of \eqref{eq:eigstates} and \eqref{eq:QXb} which proves the conjecture of \cite{Gromov:2016itr} for arbitrary $N$ previously proven only for $N=2,3$.

To render \eqref{generationofstates} more practical for usage for arbitrary $A,S$, one needs to find an effective way to construct independent operators $X_{kj}^{\alpha}$, or at least their partially symmetric combinations appearing in \eqref{generationofstates}. We hope that exploring supersymmetric generalisations of the proposed techniques will help with this matter. Indeed, suppersymmetry can "linearise" representation theory of $\gla(N)$ by reducing it to study of a chain of $\gla(1|1)$ subalgebra imbeddings \cite{Gunaydin:2017lhg}, and one can expect that the same is true for representations of Yangian as is hinted in \cite{Marboe:2016yyn,Marboe:2017dmb}. The paper \cite{Gromov:2018cvh} contains the first promising results on the properties of $\B$ in the supersymmetric case, and further study is needed.

\acknowledgments
We thank D.~Chernyak, S.~Frolov, N.~Kitanine, I.~Kostov, M.~de~Leeuw, V.~Petkova, D.~Serban,  for interesting discussions, and especially to S.~Leurent for numerous interesting discussions and participation at the initial stages of this project. The work of P.R. was partially supported by the Nordita Visiting PhD Fellowship and by the SFI grant 15/CDA/3472. The work of D.V. was partially supported by the Knut and Alice Wallenberg Foundation under grant Dnr KAW 2015.0083. D.V. is also very grateful to Institut de Math\'ematiques de Bourgogne and IPhT, C.E.A-Saclay for hospitality where a part of this work was done. We would also like to thank the reviewer for reading the manuscript and for their positive feedback.
 
\bibliographystyle{utphys}
\bibliography{References}
\end{document}